\documentclass[useAMS,usenatbib]{mn2e}
\usepackage{graphicx}
\usepackage{subfigure}
\usepackage{amsmath}
\usepackage{cancel}

\title[The lifetimes of broad bridges]{Isolating signatures of major cloud--cloud collisions II: The lifetimes of broad bridge features}

\author[T. J. Haworth et al.]
{\parbox{\textwidth}{T. J. Haworth$^{1}$\thanks{E-mail: \texttt{thaworth@ast.cam.ac.uk}},
K. Shima$^{2}$,
E. J. Tasker$^2$,
Y. Fukui$^3$,
K. Torii$^3$,
J. E. Dale$^{4}$,
\\K. Takahira$^2$
and A. Habe$^2$
}\vspace{0.4cm}\\
\parbox{\textwidth}{$^{1}$Institute of Astronomy, Madingley Rd, Cambridge, CB3 0HA, UK\\
$^{2}$ Department of Physics, Faculty of Science, Hokkaido University, Kita-ku, Sapporo 060-0810, Japan\\
$^{3}$ Department of Physics and Astrophysics, Nagoya University, Chikusa-ku, Nagoya, Aichi 464-8601, Japan\\
$^{4}$ Excellence Cluster `Universe', Boltzmannstr. 2, 85748 Garching, Germany.}}
\begin{document}

\date{Accepted ???. Received ???; in original form ???}

\pagerange{\pageref{firstpage}--\pageref{lastpage}} \pubyear{2012}

\maketitle
\label{firstpage}

\begin{abstract}
We investigate the longevity of broad bridge features in position--velocity diagrams that appear as a result of cloud--cloud collisions. Broad bridges will have a finite lifetime due to the action of feedback, conversion of gas into stars and the timescale of the collision. We make a series of analytic arguments with which to estimate these lifetimes. Our simple analytic arguments suggest that for collisions between clouds larger than $R\sim10$\,pc the lifetime of the broad bridge is more likely to be determined by the lifetime of the collision rather than the radiative or wind feedback disruption timescale. However for smaller clouds feedback becomes much more effective. This is because the radiative feedback timescale scales with the ionizing flux $N_{ly}$ as $R^{7/4}N_{ly}^{-1/4}$ so a reduction in cloud size requires a relatively large decrease in ionising photons to maintain a given timescale. We find that our analytic arguments are consistent with new synthetic observations of numerical simulations of cloud--cloud collisions (including star formation and radiative feedback). We also argue that if the number of observable broad bridges remains $\sim$ constant, then the disruption timescale must be roughly equivalent to the collision rate. If this is the case our analytic arguments also provide collision rate estimates, which we find are readily consistent with previous theoretical models at the scales they consider (clouds larger than about 10\,pc) but are much higher for smaller clouds. 
\end{abstract}


\section{Introduction}

Collisions between giant molecular clouds in galactic discs are a candidate mechanism for triggering the formation of massive stars and potentially even massive star clusters. When two clouds collide, a dense layer of material rapidly forms  which might provide the conditions required for high mass star formation. This process has been studied in numerical simulations by, for example \cite{1970ApJ...159..277S}, \cite{1992PASJ...44..203H}, \cite{1997ApJ...485..254R}, \cite{2010MNRAS.405.1431A}, \cite{2013ApJ...774L..31I}, \cite{2014ApJ...792...63T} and \cite{2015arXiv150301873W}, which show that collisions can indeed create the conditions required for high mass star formation.  \cite{2015MNRAS.446.3608D} used numerical simulations of gas evolving in a galactic potential to estimate a collision rate of about 1 per 10\,Myr for clouds larger than about 10\,pc. Similar estimates of the collision frequency for clouds larger than 10\,pc have also been obtained by \cite{2009ApJ...700..358T}, \cite{2011ApJ...730...11T} and \cite{2015ApJ...801...33T}.

Observationally, many massive stars and star clusters have now been identified as possibly resulting from cloud collisions. Two particularly compelling examples are M20 \citep{2011ApJ...738...46T} and NGC3603 \citep{2014ApJ...780...36F}. These are star clusters which appear to be at a junction between two clouds with line of sight velocities differing by approximately 20km\,s$^{-1}$. Of course it is possible that such a configuration could result from an independent star formation scenario and a chance alignment of two other clouds along the line of sight. However analysis of the $^{12}$CO\,(J=3--2, 2--1 and 1--0) lines demonstrated that the components of these two separate clouds closest to the stellar cluster are being heated by the cluster - implying that they must be spatially coincident. 

Another signature of cloud--cloud collisions is a broad bridge feature in position--velocity diagrams (hereafter p--v diagrams). This is two velocity peaks, spatially coincident but separated in velocity, with intermediate intensity gas at intermediate velocities. In \cite{2015MNRAS.450...10H} we produced synthetic p--v diagrams from a range of different simulations of molecular clouds, including cloud collisions and isolated clouds with radiative feedback. We found that indeed broad bridges appeared in our cloud collision models, but did not arise in any of the simulations of isolated clouds with radiative feedback (though in principle, a broad bridge could possibly arise given a favourable configuration of gas clouds merely coincident along the line of sight). \cite{2015MNRAS.450...10H} also compared the simulated p--v diagrams with observed broad bridge features towards M20, a site of collision according to \cite{2011ApJ...738...46T}. 

Broad bridges in p--v diagrams clearly \textit{can} arise from cloud--cloud collisions, however it is currently unclear for how long they should survive. Feedback from winds and the ionising radiation field of any OB stars that form, as well as the conversion of gas into stars and the finite lifetime of the collision will presumably limit the amount of time that broad bridge features are visible for. In this paper we combine a series of analytic arguments with new synthetic observations of cloud collision models, both with and without radiative feedback, to try and address this question. 




\section{Analytic arguments regarding  the timescale for disruption of broad bridges }

In this section we explore the time for which a broad bridge feature might survive. There are many processes that might act to remove the broad bridge such as the collision being completely braked or ending as one cloud punches through the other, radiative/mechanical feedback from massive stars and the conversion of gas into stars.

\begin{figure}
	\hspace{-0pt}
	\includegraphics[width=8.5cm]{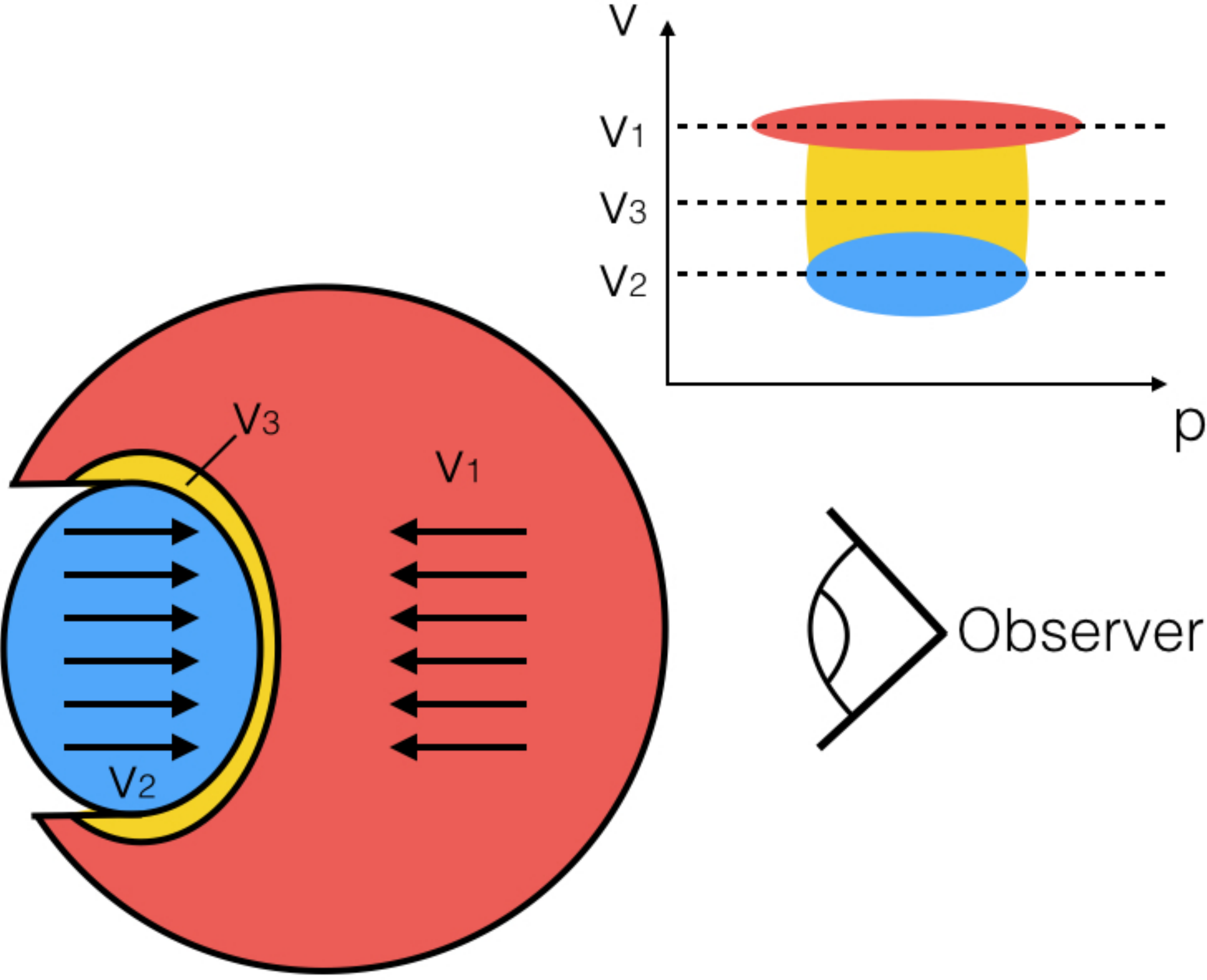}
	\caption{A schematic of a collision--observer system and a cartoon p--v diagram showing a broad bridge feature. Different colour components in the collision schematic correspond to the different colours on the p--v diagram. }
	\label{schematic}
\end{figure}

\subsection{Preamble: what is the broad bridge}
\begin{figure}
	\hspace{-20pt}
	\includegraphics[width=9.2cm]{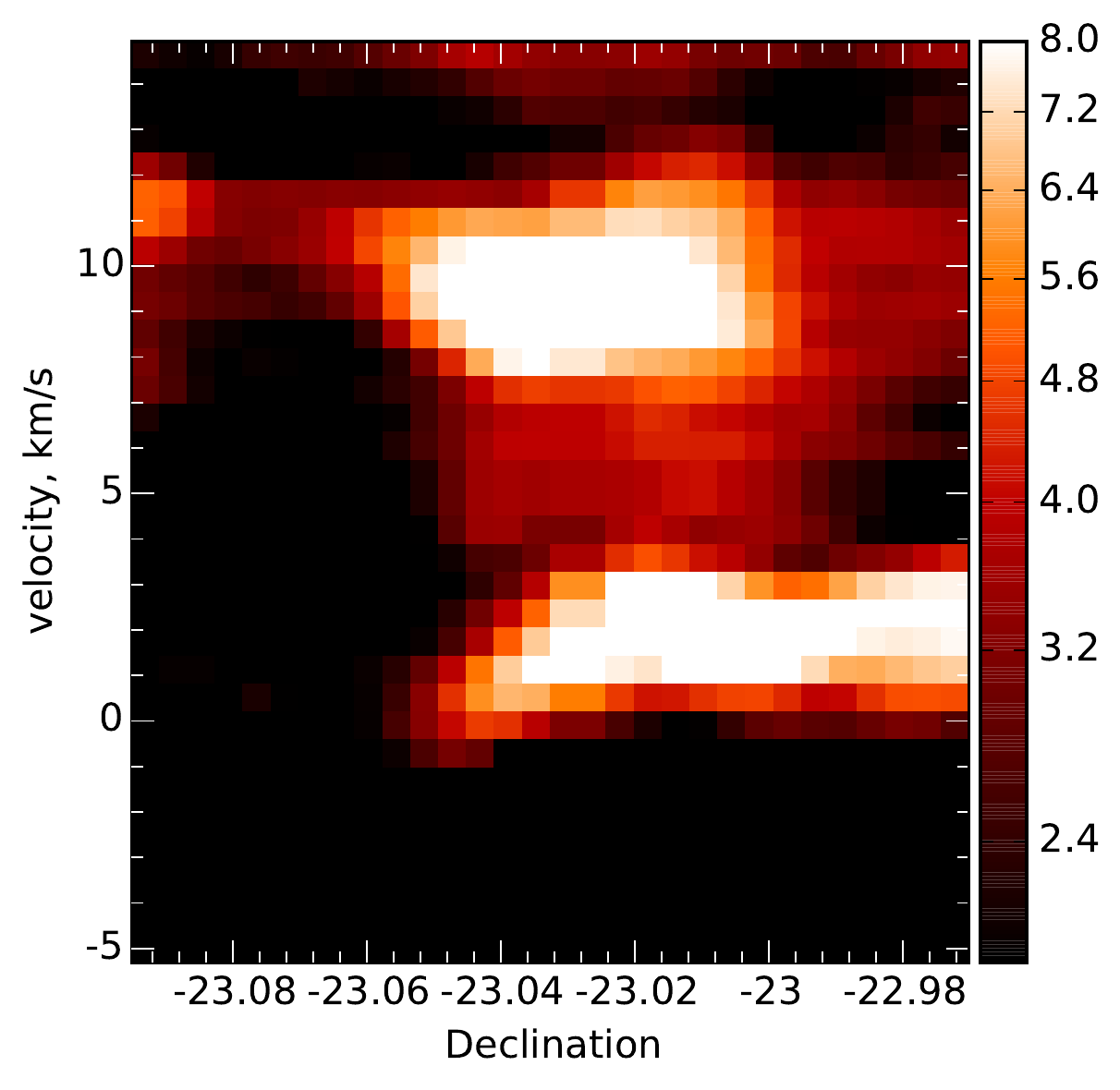}
	\caption{An example CO (J=$1\rightarrow0$)  p--v diagram of a possible cloud--cloud collision towards M20 \citep{2011ApJ...738...46T}. This p--v diagram is produced using data taken using Mopra and was presented in Haworth et al (2015).}
	\label{arrows}
\end{figure}

We begin by recapping what the broad bridge feature is in the context of a collision. Figure \ref{schematic} shows a schematic of a cloud--cloud collision, as well as a cartoon of the p--v diagram the observer would see. A smaller cloud (blue) drives into a larger cloud (red). The entirety of the smaller cloud undergoes collision quite quickly, resulting in a compressed layer that continues to move into the larger cloud. Between the bulk velocities of the compressed layer and larger cloud there is intermediate velocity gas (yellow). For the observer viewing angle in the schematic, the observer sees some red--shifted component moving away from them, a blue shifted component moving towards them and the intermediate velocity gas. 
In the p--v diagram this manifests itself as two peaks along the velocity dimension, separated by lower intensity intermediate velocity emission - the broad bridge feature. An example of a p--v diagram with a broad bridge towards in M20, taken with Mopra, is given in Figure \ref{arrows}. 

The intermediate velocity gas is replenished as long as the collision is still occurring, so in order to remove the broad bridge feature the collision either needs to end (being fully braked or with one cloud punching right through the other), or the {compressed dense layer resulting from the smaller cloud} needs to be at least partially removed.



\subsection{Details of the collision}
We consider a collision between two clouds in a scenario similar to that discussed by \cite{1992PASJ...44..203H}. One cloud of radius $R_1$ collides  with a second, larger ($R_2$), cloud at velocity $v_c$. The entirety of the small cloud has been compressed after a time approximately given by  $t_c\approx R_1/v_c$ (this is a lower limit, since it assumes that the post--collision velocity is small). We work in a frame in which the larger cloud is stationary. After $t_c$ the compressed layer is still moving with some finite bulk velocity relative to the larger (in our frame, static) cloud, meaning that the broad bridge is still observable for some subsequent timescale (to be determined below). Our considerations here are only very approximate, we do not include the effects of turbulence or instabilities that can arise in a compressive flow \citep[e.g.][]{2013MNRAS.431..710M}.

\subsection{The collision timescales}
The broad bridge will disappear if the clouds brake to the extent that the are no longer separated in velocity by more than half the sum of their turbulent velocity dispersions. If this happens then only a single feature will appear in the p--v diagram. The broad bridge may also disappear if the smaller cloud punches right through the larger. To estimate these timescales we require a description of the smaller cloud velocity as a function of time.

\begin{figure}
	\hspace{-2pt}
	\includegraphics[width=9.25cm]{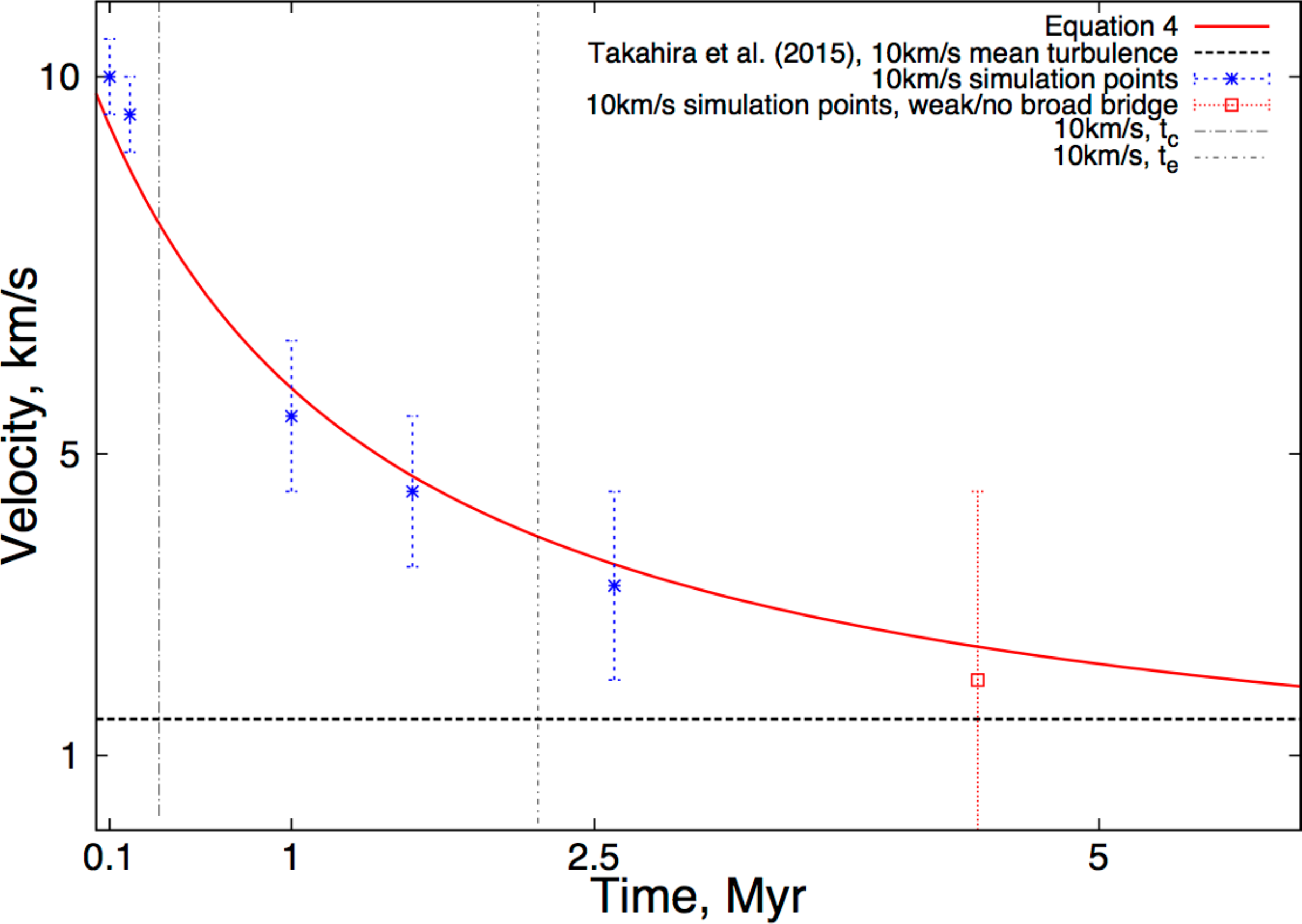}

	\vspace{5pt}
	\hspace{-2pt}	
	\includegraphics[width=9.25cm]{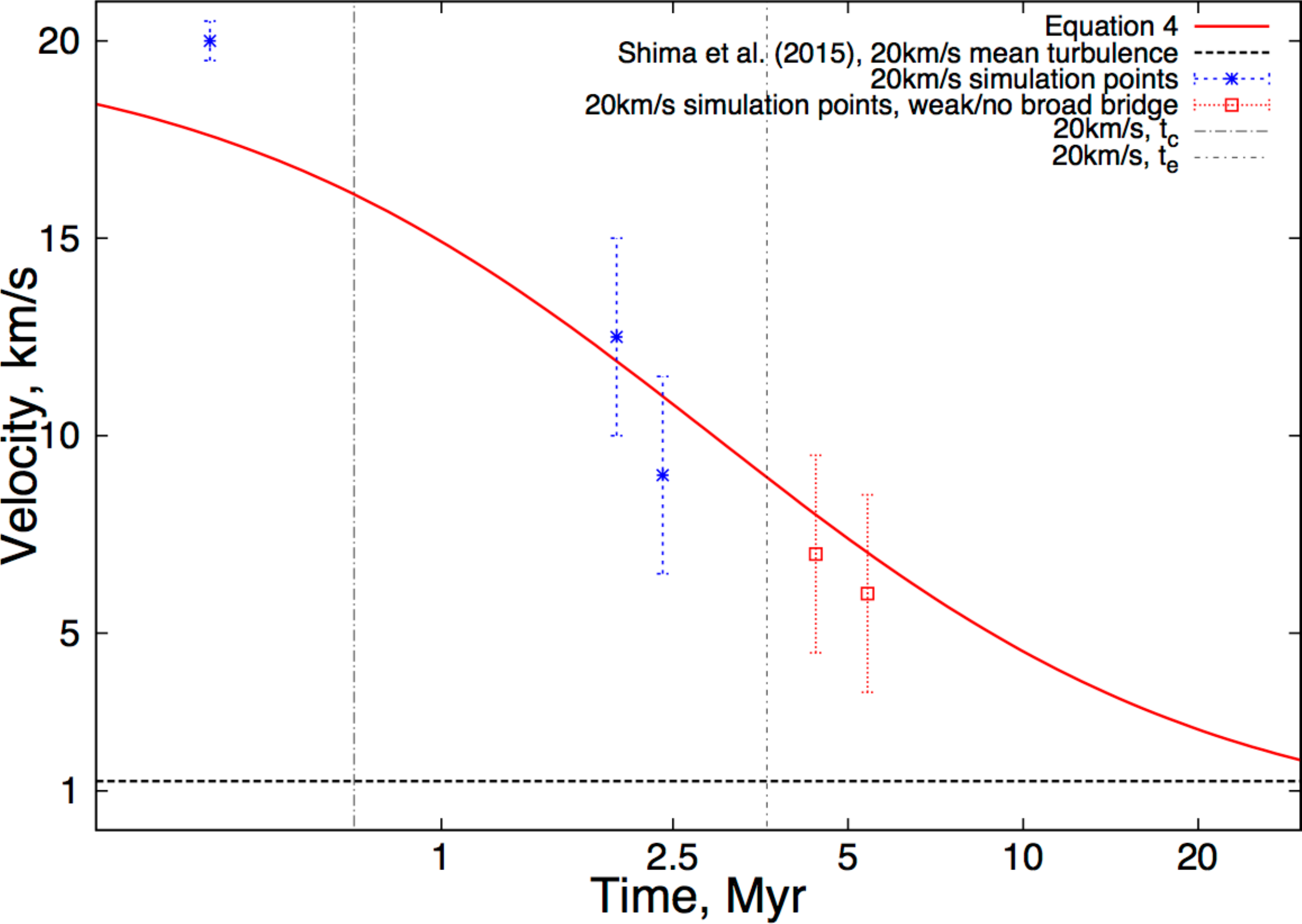}	
	\caption{The velocity of the compressed layer as a function of time estimated using momentum balance and by assuming constant deceleration. The points are extracted from p--v diagrams/line profiles from synthetic data cubes of simulations (which we will discuss in section 3). The width of the secondary peak in the broad bridge, or ease with which two distinct velocity components can be identified determines the size of the error bars. The left hand vertical line denotes the time after which the small cloud is compressed. The right hand vertical line denotes the time after which the smaller cloud has punched through the larger. }
	\label{momPlot}

\end{figure}

\subsubsection{Momentum driven motion in the limit of constant deceleration}
The smaller cloud (compressed layer) collides with the larger cloud in a frame in which the large cloud is initially static. We assume that all material from the larger cloud encountered by the compressed layer is swept up to travel with it. By conservation of momentum, this sweeping up of material decreases the velocity in the compressed layer and will lead to the collision being completely stalled after a  distance
\begin{equation}
	D_{stag} = \frac{2\rho_1R_1v_c}{\rho_2v(t)}.
\end{equation}
Assuming constant deceleration from the collision velocity $v_c$ to zero
\begin{equation}
	\frac{dv}{dR} = \frac{v_{c}}{D_{stag}} = \frac{\rho_2v(t)}{2\rho_1R_1}
\end{equation}
which yields\begin{equation}
	\frac{dv}{dt}=\frac{dv}{dR}\frac{dR}{dt} = \frac{\rho_2v(t)^2}{2\rho_1R_1}
\end{equation} 
integrating this gives us a crude estimate of the velocity of the smaller cloud relative to the larger as a function of time
\begin{equation}
	v(t) = \frac{1}{\frac{\rho_2t}{2\rho_1R_1} + \frac{1}{v_c}} = \frac{2R_1\rho_1v_c}{\rho_2 tv_c + 2\rho_1R_1}
	\label{voft}
\end{equation}
This is shown for the 20\,km\.s$^{-1}$ collision between large clouds from \cite{shimaInPrep} as well as the 10\,km\,s$^{-1}$ collision model from \cite{2014ApJ...792...63T} in Figure \ref{momPlot}. The \cite{shimaInPrep} models include sink formation and radiative feedback whereas the \cite{2014ApJ...792...63T} models do not. Included  in Figure \ref{momPlot} are also points inferred from synthetic molecular line data cubes of the simulations (these synthetic observations will be discussed in more detail below). There are three time regions denoted on each plot. The leftmost region ($t<t_{c}$) is prior to the crossing timescale of the smaller cloud. In this region we don't necessarily expect good agreement because only some small fraction of the smaller cloud is undergoing collision, which the analytic model does not account for. In the middle time region the whole small cloud is compressed into a layer travelling with a similar bulk velocity (ignoring turbulence), we therefore expect quite good agreement with our simple analytic estimate in this regime. The right hand time region is beyond the timescale that we predict the smaller cloud has exited the larger. In this region it is again difficult to estimate the velocities from p--v maps since the broad bridge is completely removed or unclear. The derived estimates in this final time zone are made using a combination of p--v maps, the raw simulation data and line profiles.

The simple analytic approximation reproduces the values seen in the simulations to within (at worst) $\sim$20 per cent and will therefore be a suitable tool for giving a first order estimate of the timescale over which the broad bridge survives due to the overall collision lifetime. Although after $t_c$ the velocities predicted by our simple relation lie within the uncertainties (which are mostly based on how broad the two peaks of the broad bridge are) the simulations typically exhibit lower velocities than the analytic model predicts. This could be due to turbulent ram pressure braking the cloud further. The analytic solution also does not include gravity, which could affect the motion. 

Equation \ref{voft} integrated until the collision velocity is less than half the sum of the two clouds turbulent velocity dispersions
\begin{equation}
 v(t)< \frac{1}{2}(\Delta v_1 + \Delta v_2),
 \end{equation}
 or until the compressed layer escapes the larger cloud (traverses $R_1 + R_2$) gives us our first broad bridge lifetime estimate. For the latter criterion we obtain an exit timescale for the smaller cloud of
\begin{equation}
	t_e = \frac{2\rho_1R_1}{\rho_2v_c}\left[\exp(\frac{\rho_2(R_1+R_2)}{2\rho_1R_1})-1\right].
	\label{texit}
\end{equation}
This is our first broad bridge lifetime estimate, we will utilise it in section \ref{applyAll}

\subsection{Radiative feedback disruption timescale}
\label{rfeedbacktime}
During the collision material is compressed into a dense layer, which continues to travel into the larger cloud, giving rise to the secondary feature in the p--v diagram. It is also  within this dense region that star formation (possibly massive star formation) will occur. However, once stars have formed their ionising radiation will result in bubbles of hot gas that might disrupt the dense layer, potentially rendering it unobservable (at least in molecular lines) once some fraction of it has been ionised. 

For a star emitting $N_{ly}$ ionising photons per second into a medium of number density $n$ a bubble of ionised gas results, with a ``Str\"{o}mgren'' radius \citep{1939ApJ....89..526S} 
\begin{equation}
	r_s = \left(\frac{3N_{ly}}{4\pi n^2 \alpha_{B}} \right)^{1/3}
	\label{stromgren}
\end{equation}
where $\alpha_{B}$ is the hydrogen case B recombination coefficient (that for recombinations in to all states other than the ground). This bubble {is hot (10$^4$\,K) and overpressured relative to the ambient medium, meaning it expands}. For our purposes, the extent of the H\,\textsc{ii} region (with sound speed $c_i$) under such so--called ``D-type'' expansion can be adequately described by the relation derived by \cite{1998ppim.book.....S}
\begin{equation}
	r(t) = r_s\left(1 + \frac{7c_it}{4r_s}\right)^{4/7}
\end{equation}
(though the interested reader is directed towards \cite{2015MNRAS.453.1324B},  for a discussion on the shortcomings and interpretation of the Spitzer solution, including how it differs from the solution derived by Hosokawa and Inutsuka, 2006). As the ionised region expands it will disrupt a larger volume of the material that makes up the broad bridge feature, until at some point it renders it undetectable.

We assume that the compressed layer has a length approximately the same as the smaller cloud diameter, $2\times R_1$ and that the width is less than the Str\"{o}mgren radius {(see appendix A)}. 
If this is the case, and we assume that the stars are centrally condensed in a cluster, then we only need to calculate the extent along the length of the layer that  is ionised as a function of time. If the broad bridge is unobservable after some fraction $f$ of the compressed layer is ionised, then we can calculate the broad bridge lifetime in the presence of radiative feedback using $fR_1 = r(t)$, giving
\begin{equation}
	t_r = \frac{4r_s}{7c_i}\left[\left(\frac{fR_1}{r_s}\right)^{7/4}-1\right]. 
	\label{t_r}
\end{equation}
We can recast $r_s$ in terms of collision variables. If the shock is isothermal, then the number density just behind the shock is
\begin{equation}
	n = \frac{\rho_2}{m_H}\left(\frac{v_L}{c_2}\right)^2
	\label{n}
\end{equation}
{where $v_L$, $\rho_2$ and $c_2$ are the dense layer velocity and the larger cloud density and sound speed respectively}.  Substituting equation \ref{n} into equation \ref{stromgren} results in
\begin{equation}
	r_s = \left(\frac{3N_{ly} m_H^2 c_2^4}{4\pi\alpha_B \rho_2^2 v_L^4}\right)^{1/3}
	\label{strom2}
\end{equation}
which combined with equation \ref{t_r} gives us a radiative feedback disruption timescale as a function of the large cloud density and collision velocity
\begin{equation}
	t_r = \frac{4}{7c_i}\left(\frac{3N_{ly} m_H^2 c_2^4}{4\pi\alpha_B \rho_2^2 v_L^4}\right)^{1/3}\left[\left(\frac{fR_1}{\left(\frac{3N_{ly} m_H^2 c_2^4}{4\pi\alpha_B \rho_2^2 v_L^4}\right)^{1/3}}\right)^{7/4}-1\right]. 
	\label{big_t_r}
\end{equation}
{At the first level of approximation, we might set the dense layer velocity equal to the collision velocity ($v_L=v_c$). Indeed if these equations are applied to observations, we suggest just using the observed collision velocity.  However, for our discussion here it is more accurate to use the mean velocity over the lifetime of the broad bridge feature. To estimate the mean velocity, we first calculate an initial estimate of $t_r$ using the velocity at the onset of collision $v_c$. We then calculate the mean velocity over this initial $t_r$  timescale estimate using equation \ref{voft}. Lastly, we iterate over $t_r$ and $v_L$ until the fractional change in mean velocity is less than 1 per cent. Typical mean velocities over the radiative timescale are around 40 to 60 per cent of the initial collision velocity. }

\begin{figure}
	\hspace{-10pt}
	\includegraphics[width=9cm]{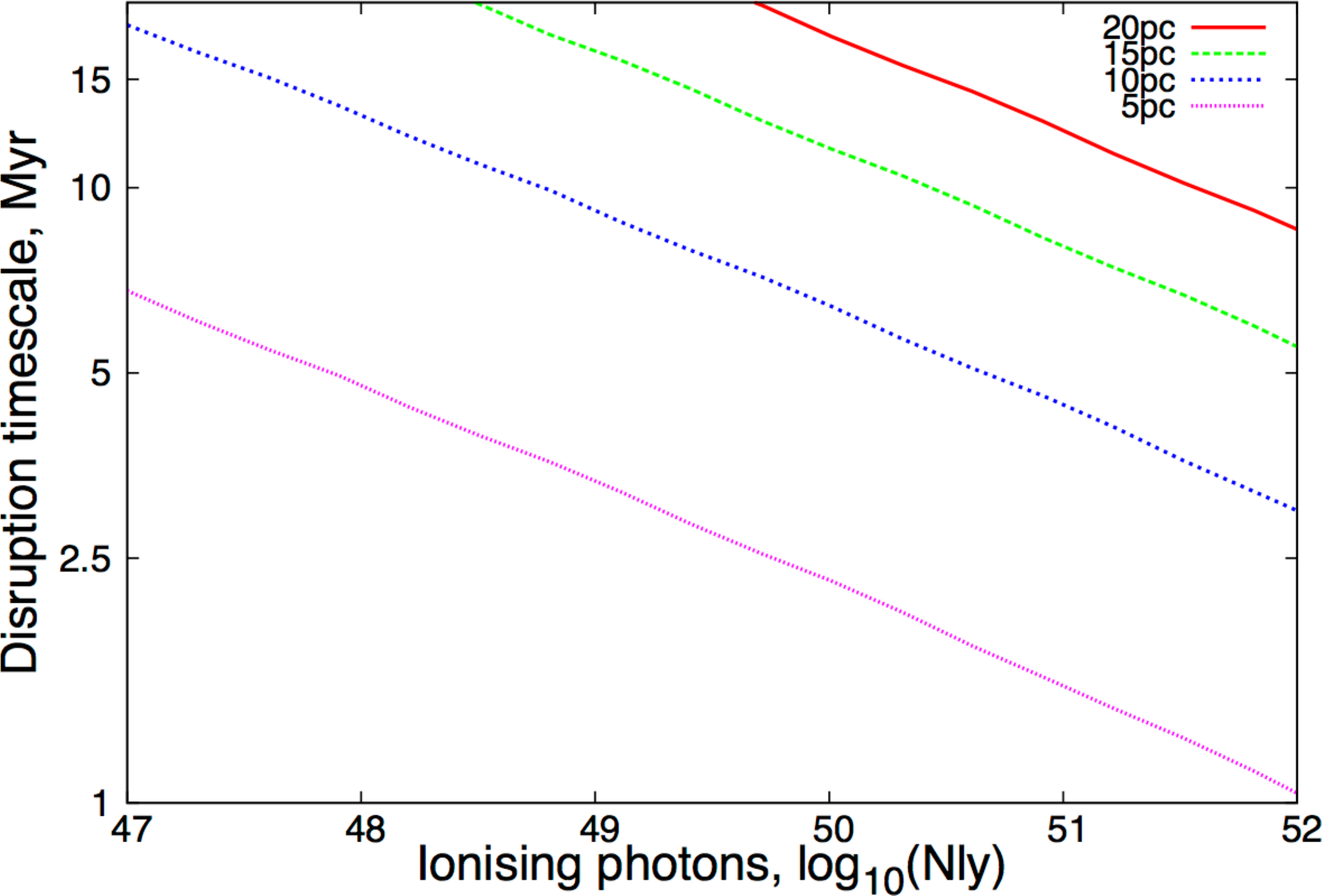}
	
	\hspace{-10pt}	
	\includegraphics[width=9cm]{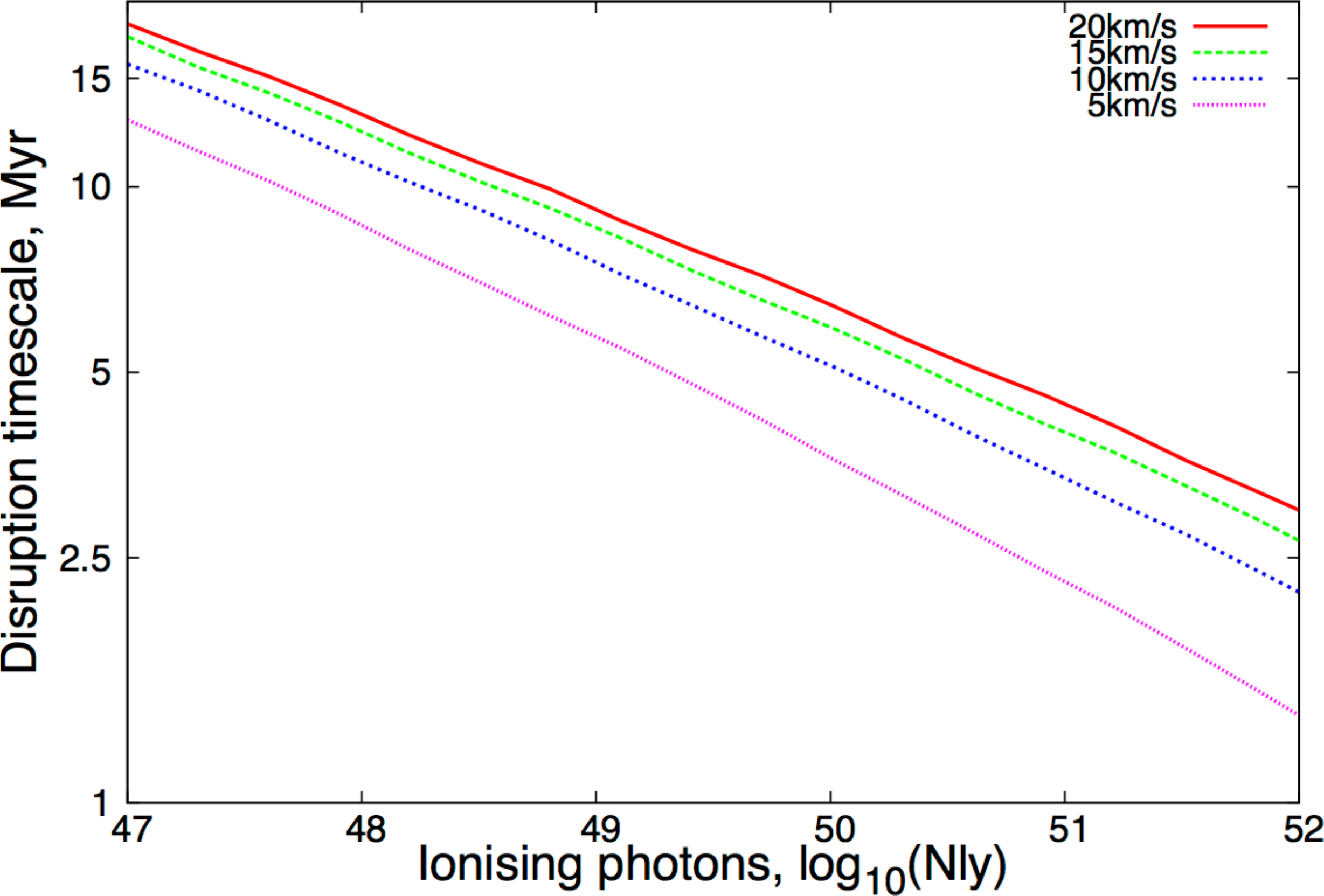}	
	\caption{The timescale for a broad bridge to be disrupted by radiative feedback as a function of ionising flux. The upper panel is for a 20\,km\,s$^{-1}$ collision for different cloud radii. The lower panel is for a 10\,pc cloud and different collision velocities. These plots assume that both colliding clouds have a density of 100\,m$_{\textrm{H}}$\,cm$^{-3}$.}
	\label{collplot}
\end{figure}

We plot the disruption timescale as a function of ionising flux for different cloud/collision parameters in Figure \ref{collplot}, each assuming that the clouds have a density of 100\,m$_H$\,cm$^{-3}$. {We assume that the ionised and neutral gases have temperatures (sound speeds) of 10$^4$ and 10\,K (11.5 and 0.19\,km/s) respectively}. {We also assume that the critical fraction of the cloud length that needs to be ionised is $f=0.5$.} Different lines {in Figure \ref{collplot}} correspond to  different smaller cloud radii and collision velocities in the upper and lower panels respectively. Since the radiative disruption timescale scales as $R_1^{7/4}$ but only $N_{ly}^{-1/4}$, a reduction of the cloud size by a factor of 10 requires a reduction in the ionizing luminosity by a factor of $10^7$ to maintain the same disruption timescale. We therefore expect broad bridges to be disrupted more readily {by radiative feedback} in collision between smaller clouds, even if the ionizing luminosity is substantially lower.  We will compare these relations with simulations in section \ref{applyAll}.

\subsection{Stellar wind feedback disruption timescale}
We can estimate the effect of stellar wind feedback using a similar approach to that for radiative feedback. That is we can compare the size of a wind blown bubble with the size of the compressed layer. We base our treatment of the expanding wind-blown region on the expansion in a cold uniform medium presented by \cite{2015ApJ...798...32N}, which in turn is based on \cite{1975ApJ...198..575S} and \cite{1999isw..book.....L}. In this model the bubble radius is given by
\begin{equation}
	r_w(t) = 0.83\dot{M_w}^{1/4}v_w^{1/4}\rho_s^{-1/4}t^{1/2}
	\label{rwind}
\end{equation}
where $\dot{M_w}$, $v_w$ and $\rho_s$ are the wind mass loss rate, velocity and slab density {(all in cgs units)} respectively. 
Comparing equation \ref{rwind} with some critical fraction of the layer size $fR_1$, and assuming an isothermal shock, gives a time at which disruption of the broad bridge due to winds takes place of
\begin{equation}
	t_w = \frac{f^2R_1^2 \rho_2^{1/2} v_L}{0.6889 \dot{M_w}^{1/2}v_w^{1/2} c_2}.
	\label{t_w}
\end{equation}

Note that this assumes that the total wind luminosity goes in to disrupting the broad bridge.


\begin{figure}
	\hspace{-10pt}
	\includegraphics[width=9cm]{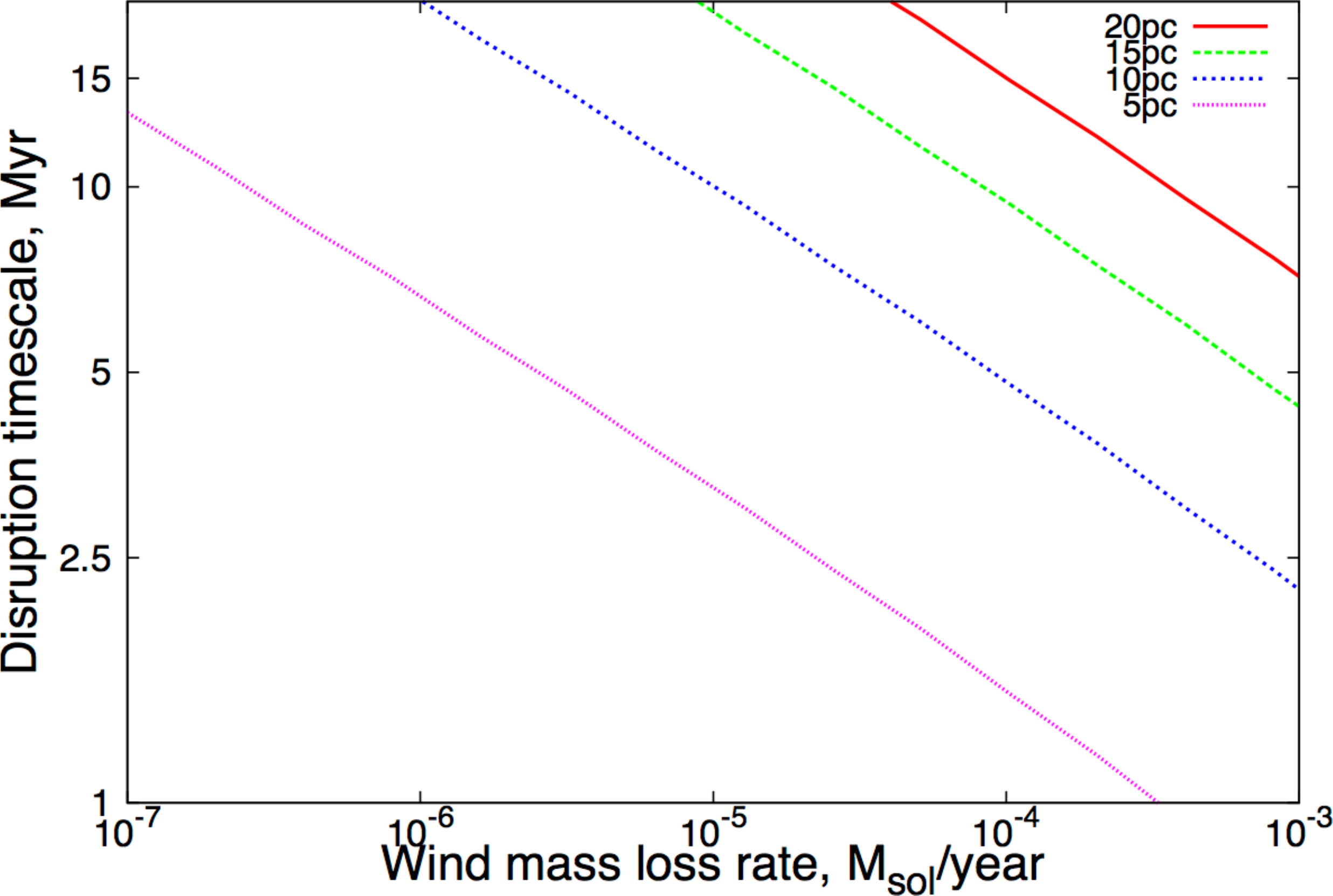}

	\hspace{-10pt}
	\includegraphics[width=9cm]{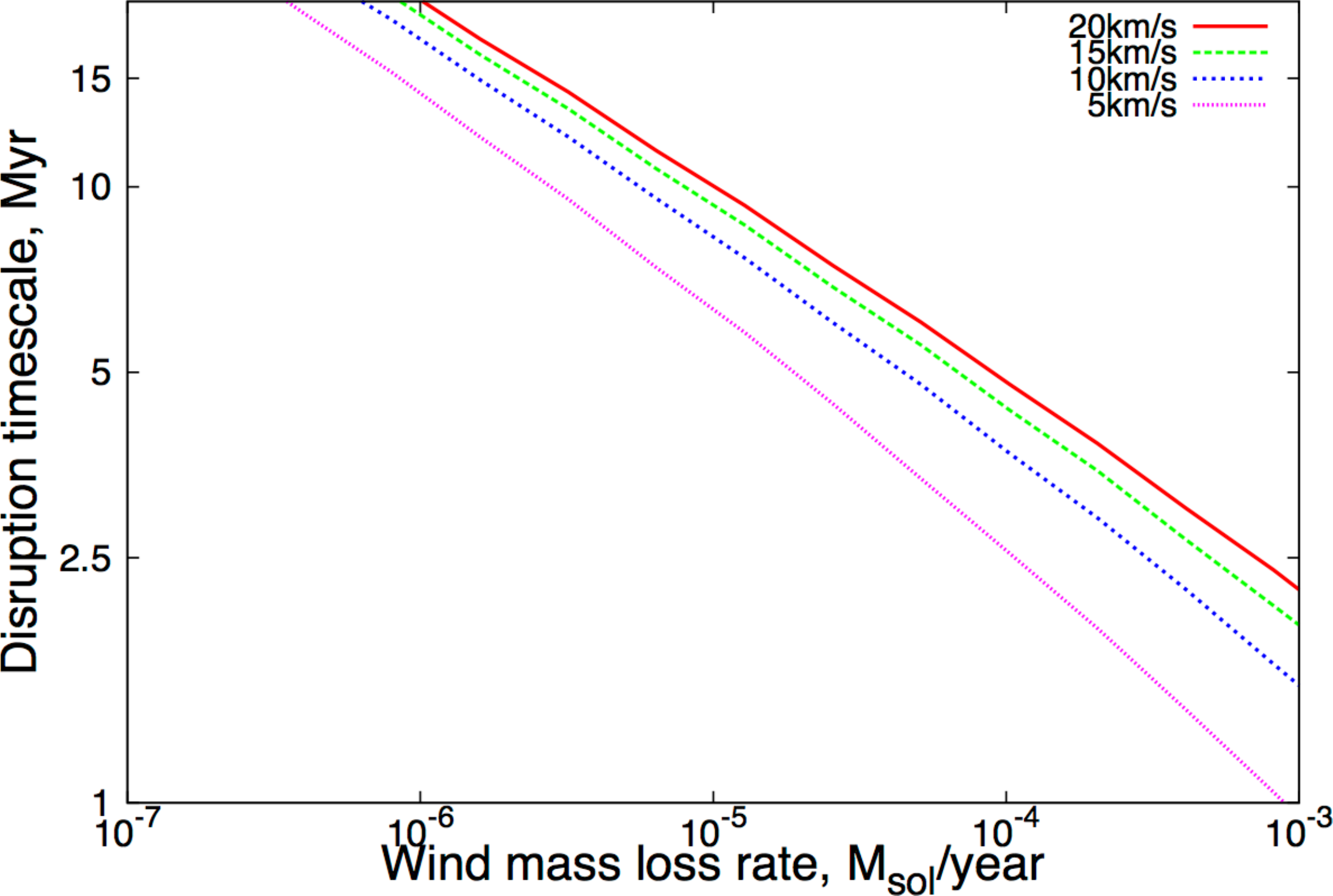}	
	\caption{The timescale for a broad bridge to be disrupted by stellar wind feedback as a function of wind luminosity for different cloud radii (at 20km\,s$^{-1}$ collision, top panel) and  collision velocities (for a cloud of 10\,pc radius, lower panel). The plots assume that both clouds have a density of 100\,m$_{\textrm{H}}$\,cm$^{-3}$.}
	\label{windcollplot}
\end{figure}

Plots of the broad bridge disruption timescale as a function of wind luminosity for a variety of cloud sizes and collision velocities are given in Figure \ref{windcollplot}. Note that in these plots we assume a wind velocity of 2000\,km\,s$^{-1}$ and a gas density of 100\,m$_{\textrm{H}}$\,cm$^{-3}$.  {We again calculate the mean dense layer velocity using the iterative procedure discussed in section \ref{rfeedbacktime}.}

By comparing equation \ref{big_t_r} with equation \ref{t_w} for systems where  $(fR_1/r_s)^{7/4} \gg 1 $ our radiative feedback timescale dominates over that for winds when
\begin{equation}
	\left(\frac{\dot{M_w}}{10^{-6}M_\odot\,\textrm{yr}^{-1}}\right) < 1.7 \left(\frac{R_1}{\textrm{pc}}\right)^{1/2}\left(\frac{N_{ly}}{10^{48}\,\textrm{s}^{-1}}\right)^{1/2}
\end{equation}
where our assumed values are that the wind velocity is 2000\,km\,s$^{-1}$, the ionised and neutral gas temperatures are 10$^4$ and 10\,K respectively, $f=0.5$ and case B recombination coefficient of $2.7\times10^{-13}$\,cm$^3$\,s$^{-1}$. 
According to previous numerical simulations by \cite{2013MNRAS.436.3430D} and  \cite{2014MNRAS.442..694D} radiative feedback is typically expected to dominate over winds.


\subsection{The star formation disruption timescale}
The gas mass depletion through star formation is roughly given by the time integral of the average star formation rate $\dot{M}$. The broad bridge will be disrupted when some fraction $F$ (which is probably different from the fraction of the length $f$ discussed in the feedback timescales) of the dense layer mass $M_{l}$  is converted into stars
\begin{equation}
	\int \dot{M}dt = FM_{l}.
\end{equation}
then the timescale for star formation to remove the broad bridge is
\begin{equation}
	t_{sf} = FM_l/\dot{M}.
	\label{t_sf}
\end{equation}
Unfortunately  $t_0$, $F$, $M_l$ and $\dot{M}$ are all rather uncertain, so a more useful relation is difficult to derive. In subsequent comparison with numerical models in this paper we therefore extract the values of $t_0$ and $\dot{M}$ from the simulations. We assume that $M_l$ is the smaller cloud mass and $F$ is 0.5. This is the most uncertain timescale in this paper.

\subsection{Weaknesses of the analytic arguments}
Our arguments made above are purely intended to provide rough lifetime estimates and to build an understanding of how feedback might disrupt the broad bridge feature. They may also assist in a basic interpretation of real observations. However, they have a number of shortcomings
\begin{enumerate}
	\item{Simulations by \cite{2011MNRAS.414..321D} find that radiative feedback can be ineffective if the H\,\textsc{ii} region is substantially matter bounded. That is, if photons are able to stream freely from the ionised bubble.  This would make radiative feedback much less effective than we predict and so will increase $t_r$.}
	\item{Regarding both winds and ionising radiation, we assume that the sources responsible for feedback are co--located. In reality they will be distributed throughout the colliding layer.}
	\item{At present there is no consideration of supernovae, which might also be effective at removing broad bridges
	after about 5\,Myr.}
	\item{We do not account for the effects of turbulence and assume constant deceleration in our collision timescale estimate. }
	\item{We assume that the collision is isothermal. In reality the compressed layer density will {likely larger than that given by the isothermal shock condition (equation \ref{n}), at least in the absence of turbulence}.}
\end{enumerate}


\section{Numerical comparison}
\label{applyAll}

We now study the evolution of the broad bridge feature in two numerical models and compare with our expectations from the analytic expressions. One of the models is  without any form of feedback or sink formation,  the second includes radiative feedback and sink formation. Neither include feedback from winds. 

\subsection{Numerical method}

\subsubsection{Hydrodynamic models}
The hydrodynamic models that we post process are simulations of cloud--cloud collisions by \cite{2014ApJ...792...63T} and Shima et al (in prep). These models were both run using the grid based radiation hydrodynamics code \textsc{enzo}  \citep{Enzo}. They both consider head--on collisions between two non--identical, turbulent clouds. 

The collision with feedback and sink particles is the 20\,km\,s$^{-1}$ collision model from Shima et al (in prep). A summary of the cloud parameters is given in {Table} \ref{ShimaModel}. Once sink particles form in this simulation they are assumed to have an ionizing luminosity of $3\times10^{46}$photons\,s$^{-1}$\,M$_{\odot}^{-1}$. In this model both clouds are travelling in opposite directions at 10\,km\,s$^{-1}$. We consider snapshots at 0.4, 2, 2.4, 4.4 and 5.4\,Myr after the onset of collision. {At 2.4\,Myr there is approximately 100\,M$_{\odot}$ in sink particles, implying an ionizing flux of  $3\times10^{48}$photons\,s$^{-1}$, this is the value we use in our radiative feedback disruption timescale estimate.}

The collision without feedback or sink particles (i.e. without star formation) is the 10\,km\,s$^{-1}$ collision model from \cite{2014ApJ...792...63T}. A summary of the clouds parameters is given in {Table} \ref{TakahiraModel}. In this model the collision is in such a frame that the larger cloud is stationary. We consider snapshots at 0.1, 0.2, 1, 1.6, 2.6 and 4.4\,Myr after the onset of the collision.

\begin{table}
 \centering
  \caption{Summary of the parameters of the feedback model from Shima et al (in prep).}
  \label{ShimaModel}
  \begin{tabular}{@{}l c c @{}}
  \hline
   Parameter & Value & Description \\
  \hline   
   $R_1$ & 14.45pc & Smaller cloud radius \\  
   $R_2$ & 27.93pc & Larger cloud radius \\  
   $\rho_1$ & 24.4\,m$_H$\,cm$^{-3}$  & Smaller cloud density \\  
   $\rho_2$ & 11.79\,m$_H$\,cm$^{-3}$ & Larger cloud density \\  
   $v_c$ & 20\,km\,s$^{-1}$ & Collision velocity\\   
   $\Delta v_1$ & 1\,km\,s$^{-1}$ & Smaller cloud turbulent velocity\\
   $\Delta v_2$ & 1.49\,km\,s$^{-1}$ & Larger cloud turbulent velocity\\
   $Q$ & $3\times10^{46}$s$^{-1}$\,M$_{\odot}^{-1}$ & {Ionizing flux per solar mass}\\   
   $Q_{tot}$ & $3\times10^{48}$s$^{-1}$ & Total ionizing flux after 2.4\,Myr\\
   & & of collision\\
\hline
\end{tabular}
\end{table}

\begin{table}
 \centering
  \caption{Summary of the parameters of the feedback model from Takahira et al (2014).}
  \label{TakahiraModel}
  \begin{tabular}{@{}l c c @{}}
  \hline
   Parameter & Value & Description \\
  \hline   
   $R_1$ & 3.5pc & Smaller cloud radius \\  
   $R_2$ & 7.2pc & Larger cloud radius \\  
   $\rho_1$ & 47.4\,m$_H$\,cm$^{-3}$  & Smaller cloud density \\  
   $\rho_2$ &25.3\,m$_H$\,cm$^{-3}$ & Larger cloud density \\  
   $v_c$ & 10\,km\,s$^{-1}$ & Collision velocity\\      
   $\Delta v_1$ & 1.25\,km\,s$^{-1}$ & Smaller cloud turbulent velocity\\
   $\Delta v_2$ & 1.71\,km\,s$^{-1}$ & larger cloud turbulent velocity\\
\hline
\end{tabular}
\end{table}

\begin{figure}
	\hspace{-20pt}
	\includegraphics[width=9.5cm]{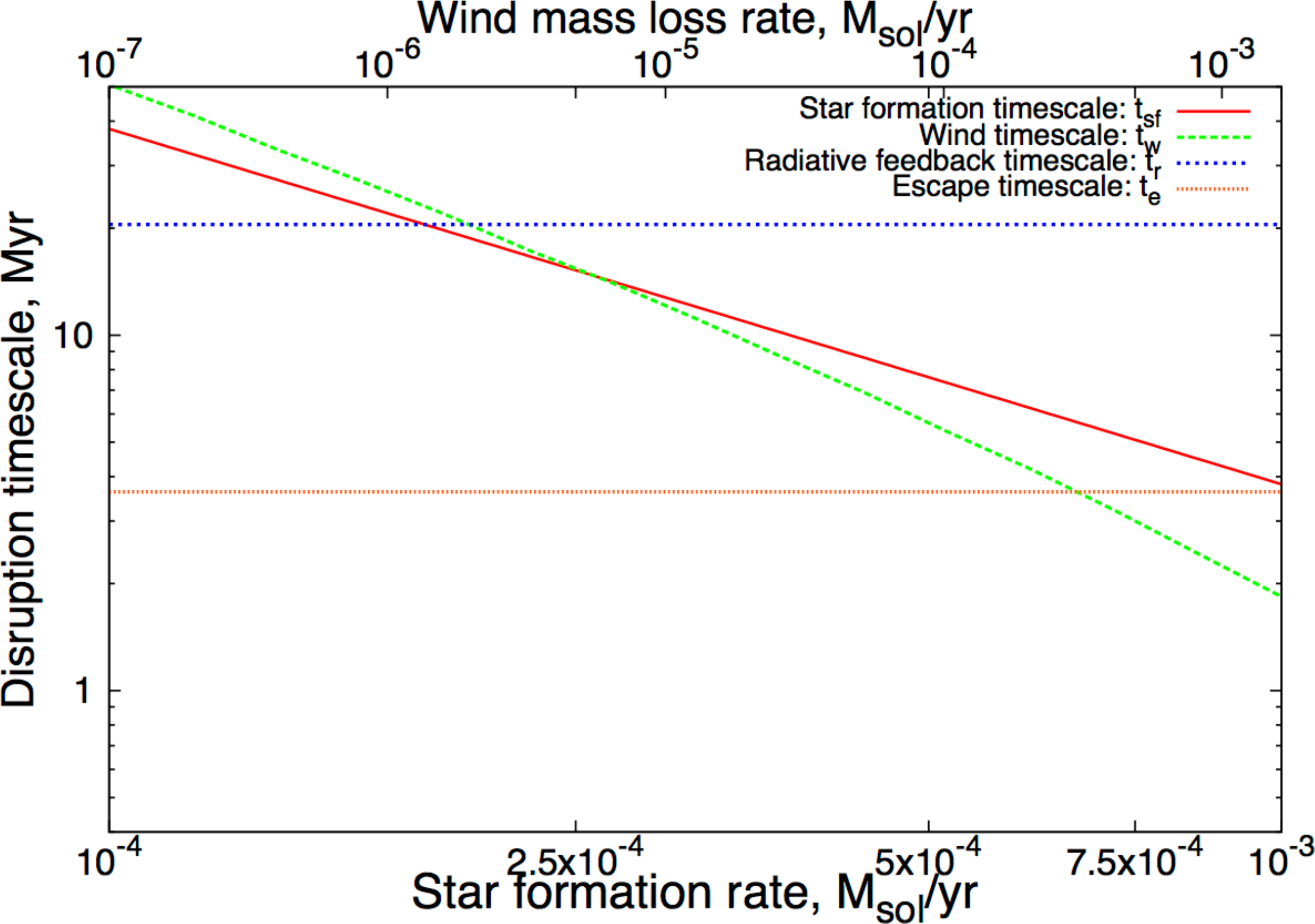}
	\caption{Various timescales for associated with the broad bridge for the 20\,km\,s$^{-1}$ collision model from Shima et al (in prep). $t_{sf}$, $t_r$ and $t_w$ are the time it takes for star formation, radiative feedback and wind feedback to quench the broad bridge. $t_e$ is the time it takes for the small cloud to punch through the larger cloud. }
	\label{allplot}
\end{figure}

\subsubsection{Radiative transfer}

As for our original synthetic observations in \cite{2015MNRAS.450...10H} we use the \textsc{torus} radiation transport and hydrodynamics code to post-process simulations in this paper \citep{2000MNRAS.315..722H, 2004MNRAS.351.1134K, 2012MNRAS.420..562H}. In particular we use the molecular line transfer algorithm discussed by \cite{2010MNRAS.407..986R}, \cite{2013MNRAS.431.3470H}. \textsc{torus} is capable of non-LTE statistical equilibrium calculations, which were used in \cite{2015MNRAS.450...10H}, however since we are prostprocessing a relatively large number of snapshots in order to study the time evolution of broad bridges we assume LTE when calculating the molecular level populations in this paper. This does not alter the qualitative p--v diagram (we are only studying the morphological evolution) and substantially reduces our computational expense. With the level populations calculated, a synthetic data cube is generated using ray tracing  \citep{2010MNRAS.407..986R}. {This data cube} is manipulated using the \textsc{starlink} image analysis tool \textsc{gaia} to produce p--v maps. In this paper we produce synthetic $^{12}$CO (J=$1\rightarrow0$) data assuming an abundance relative to hydrogen of  $8\times10^{-5}$. 

\subsection{Results of synthetic observations}

 \begin{figure*}
	\hspace{-10pt}
	\includegraphics[width=7.2cm]{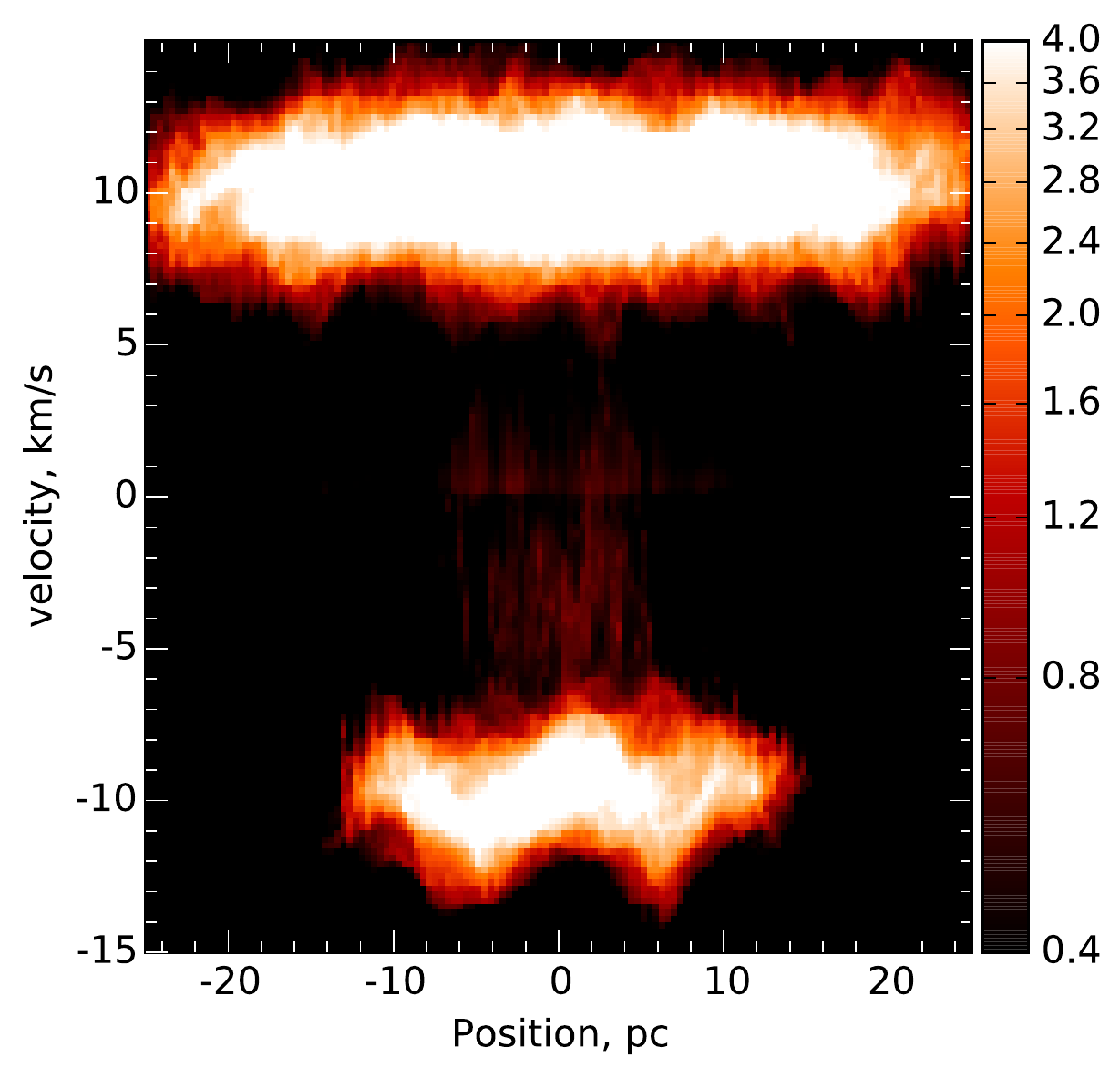}
	\includegraphics[width=7.2cm]{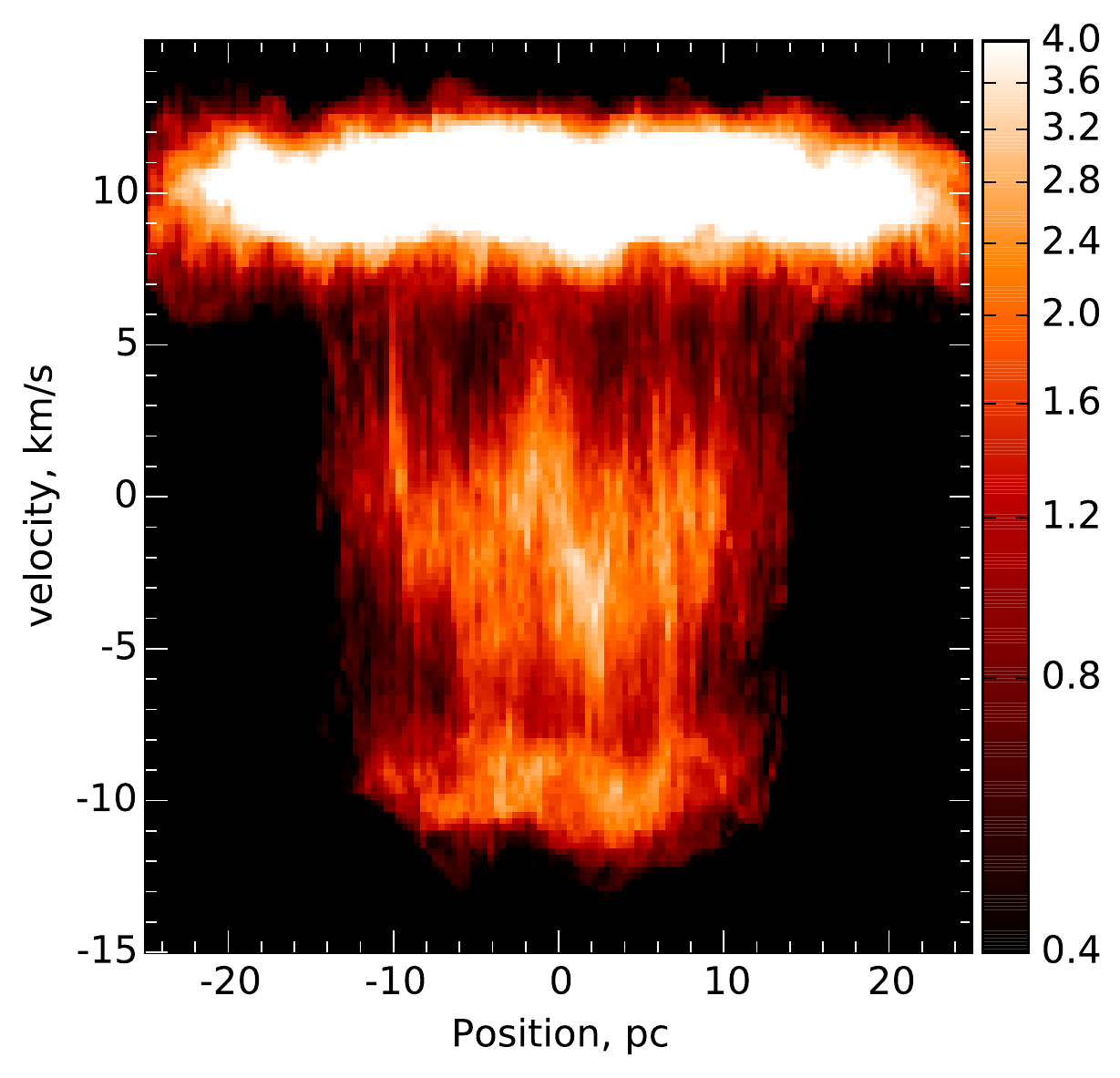}

	\hspace{-10pt}
	\includegraphics[width=7.2cm]{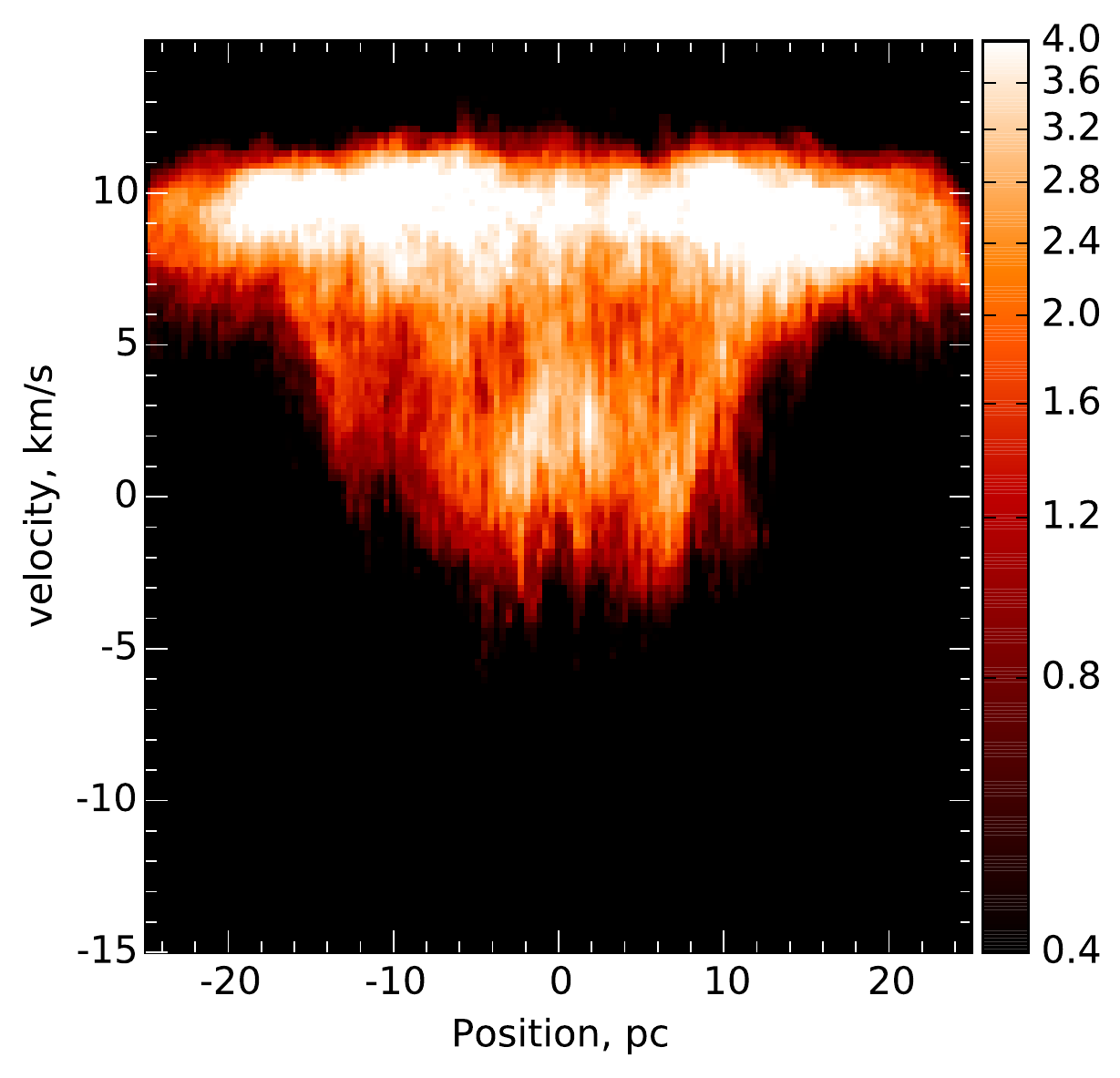}
	\includegraphics[width=7.2cm]{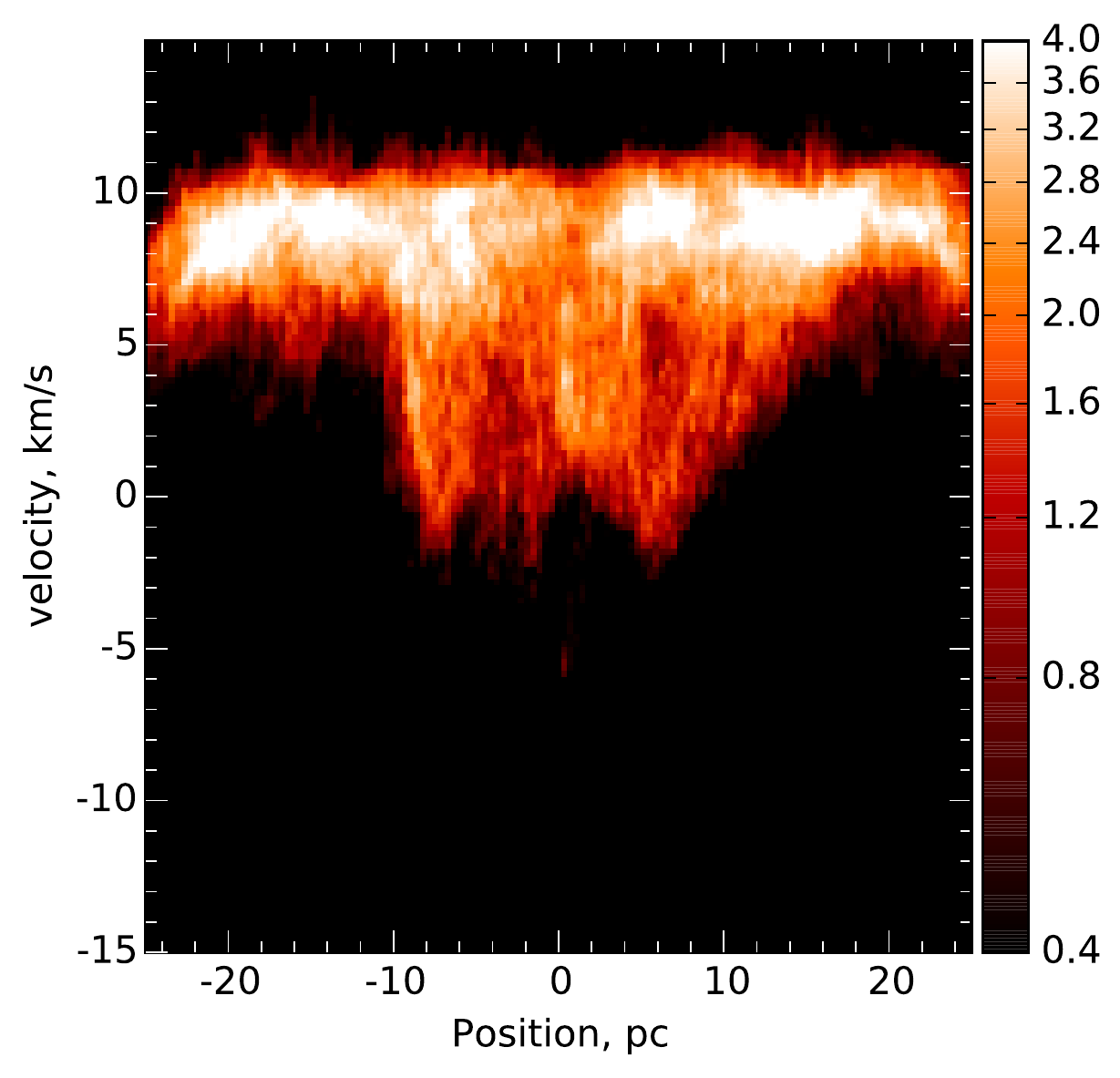}	
	\caption{$^{12}$CO (J=$1\rightarrow0$) synthetic p--v diagrams of snapshots from the 20\,km\,s$^{-1}$ collision model from Shima et al (in prep). The panels are for snapshots at 0.4 (top left), 2 (top right), 2.4 (bottom left)  and 4.4\,Myr (bottom right).}
	\label{BB20}
\end{figure*}

\begin{figure}
	\hspace{-5pt}
	\includegraphics[width=9.5cm]{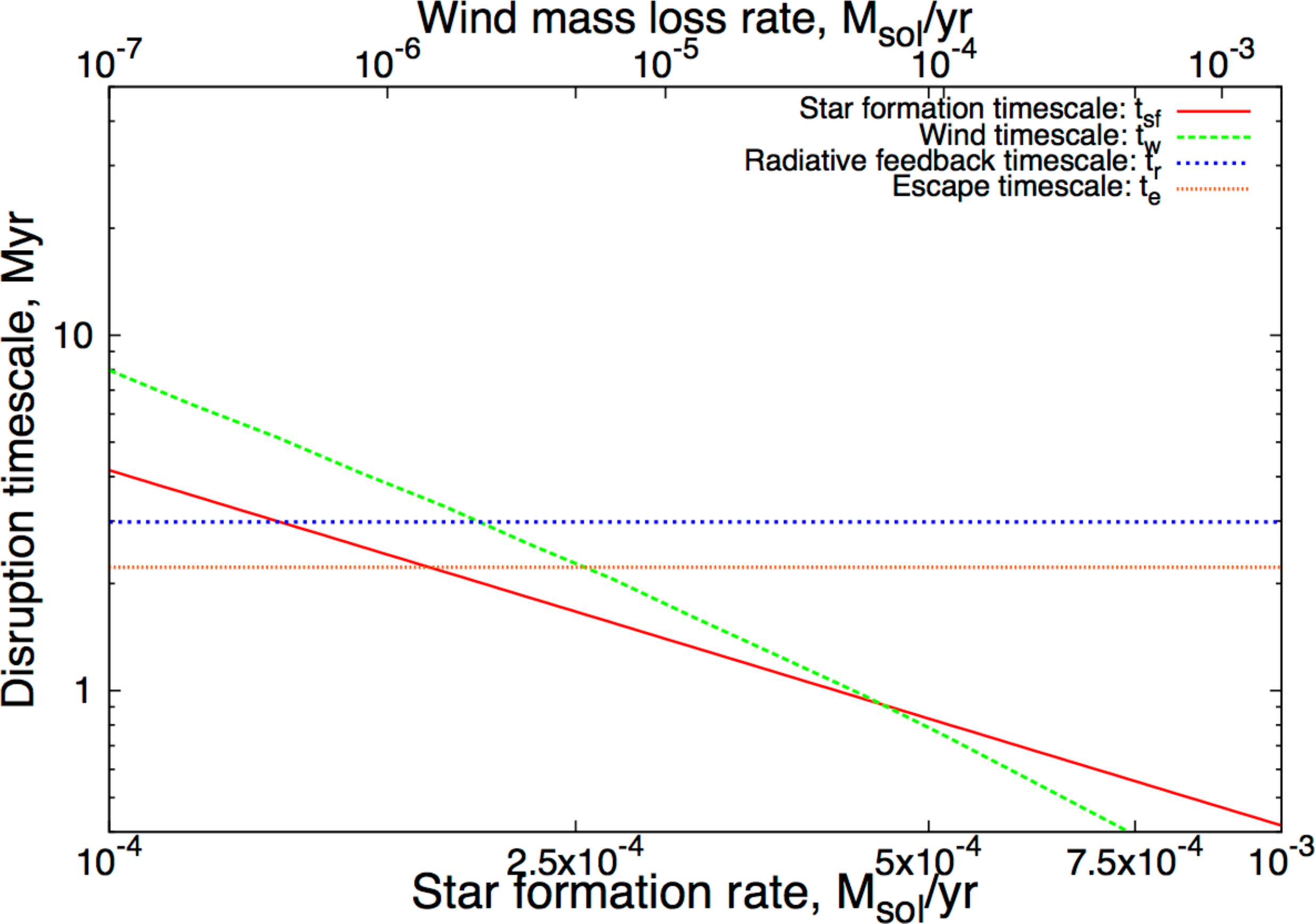}
	\caption{Various timescales for associated with the broad bridge for the 10\,km\,s$^{-1}$ collision model from Takahira et al (2014). $t_{sf}$, $t_r$ and $t_w$ are the time it takes for star formation, radiative feedback and wind feedback to quench the broad bridge. $t_e$ is the time it takes for the small cloud to punch through the larger cloud. }
	\label{allplot2}
\end{figure}

\subsubsection{20km\,s$^{-1}$ collision model from Shima et al (in prep)}
We first apply our analytic arguments to a 20\,km\,s$^{-1}$ collision between two large clouds studied by Shima et al (in prep). The parameters of this model are given in Table \ref{ShimaModel}. In Figure \ref{allplot} we plot the broad bridge disruption timescales based on the exit timescale (equation \ref{texit}),  wind/radiative feedback (equations \ref{t_r} and \ref{t_w}) and  and star formation rate (equation \ref{t_sf}) for this model. We allow the star formation rate and wind mass loss rate to vary and use the known ionizing flux of stars in the simulation at 2.4\,Myr {($3\times$10$^{48}$\,photons\,s$^{-1}$)}.  From these disruption timescales we expect the broad bridge to be first disrupted after about 3.5\,Myr because the smaller cloud has exited the larger (see the {red small--dotted} horizontal line in Figure \ref{allplot}). High star formation rates or wind mass loss rates are required to disrupt the broad bridge on a comparable timescale. We expect that radiative feedback would require about 11\,Myr to clear the broad bridge.

Figure \ref{BB20} shows the p--v diagram for this model at 0.4, 2, 2.4 and 4.4\,Myr after the onset of collision.  In the upper left panel, the clouds have only just begun colliding. This snapshot is at a time earlier than the timescale for crossing the small cloud at the collision velocity. At this stage in the collision the p--v diagram just looks like clouds coincident along the line of sight (though there is some very weak intermediate velocity emission). As the collision progresses the broad bridge forms, getting harder to identify as the two intensity peaks approach one another and as the smaller cloud's signature becomes more disordered. By 4.4\,Myr (bottom right panel) the broad bridge is obviously not present, consistent with the result expected from our simple analytic estimates. Although the broad bridge is not observable,  there is disordered emission from gas over a 10\,km\,s$^{-1}$ velocity range at this time. By 4.4\,Myr the larger cloud's feature has also started to be significantly disrupted. {The larger feature also appears to decrease in velocity dispersion with time (it's thickness along the velocity axis decreases). This is due to dissipation of the turbulence, either by collisions or dilution in the wider ISM.  }

Our analytic arguments give a broad bridge lifetime consistent with the observations to within at least 1\,Myr in this case. According to these arguments the broad bridge lifetime in this scenario is determined by the collision lifetime, rather than feedback. Though it is likely that conversion of gas into stars, feedback and the collision lifetime do all cumulatively contribute, meaning that the collision lifetime estimate is likely an upper limit on the broad bridge lifetime. 


\begin{figure*}
	\hspace{-10pt}
	\includegraphics[width=7.2cm]{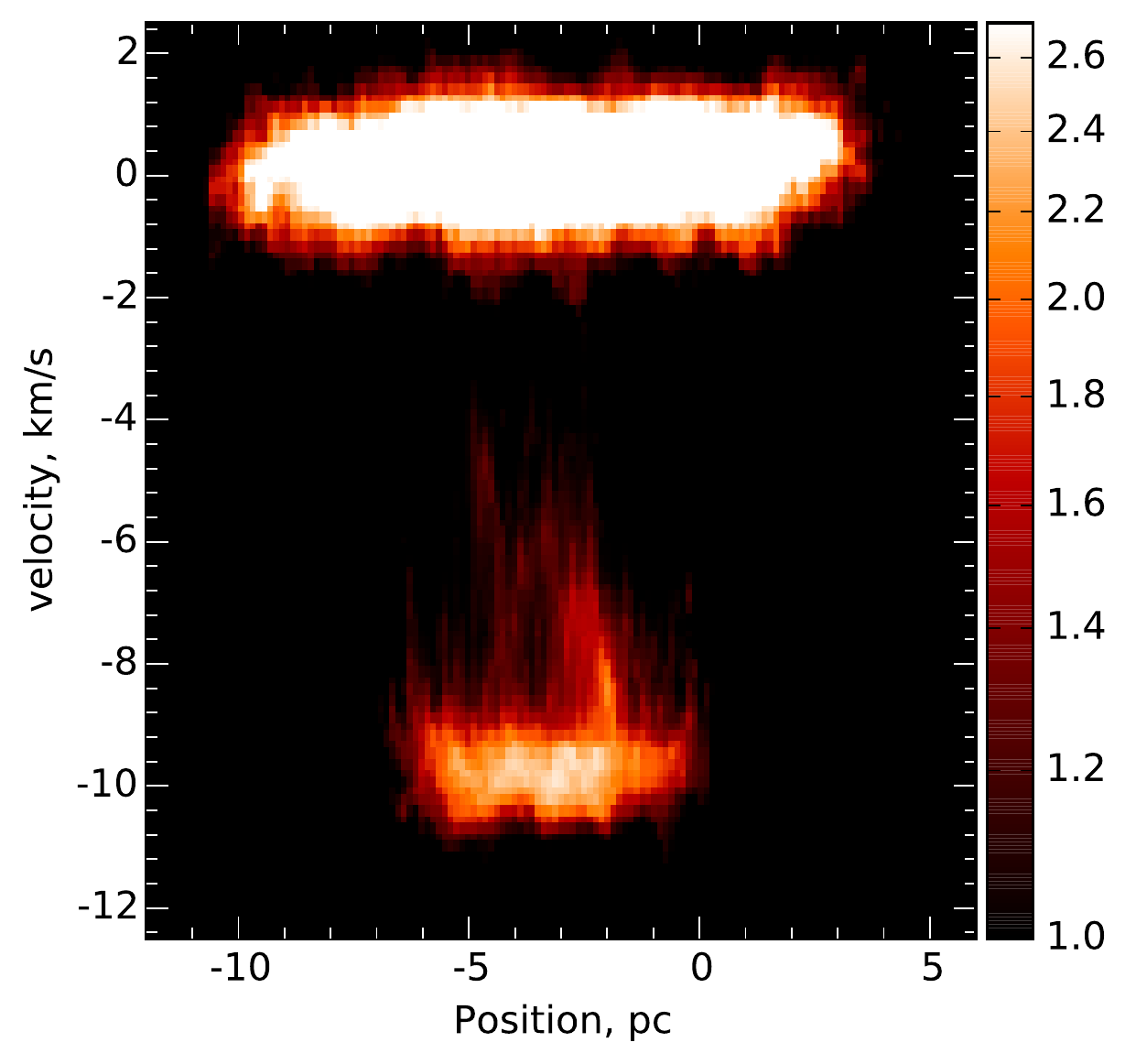}
	\includegraphics[width=7.2cm]{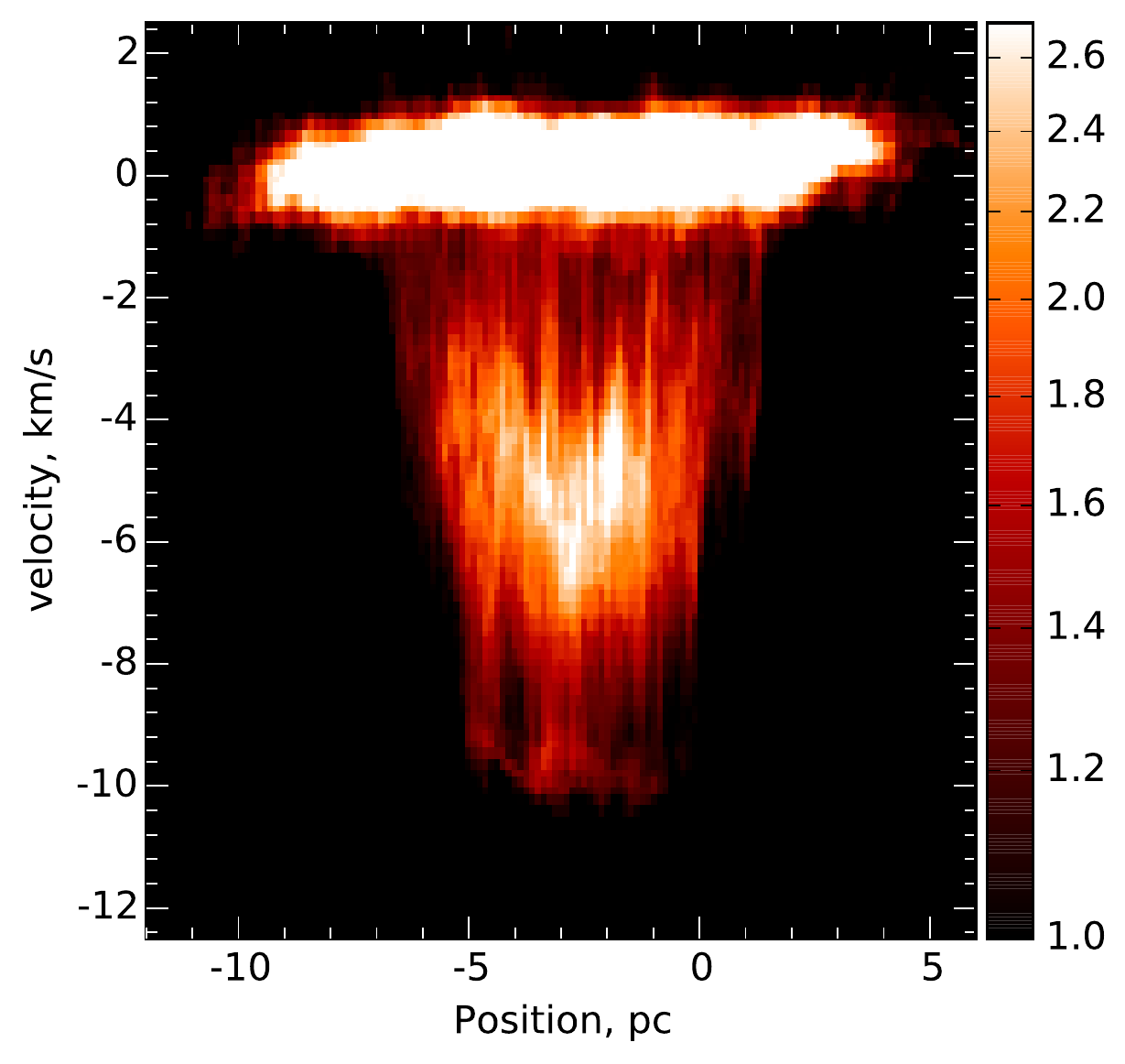}

	\hspace{-10pt}
	\includegraphics[width=7.2cm]{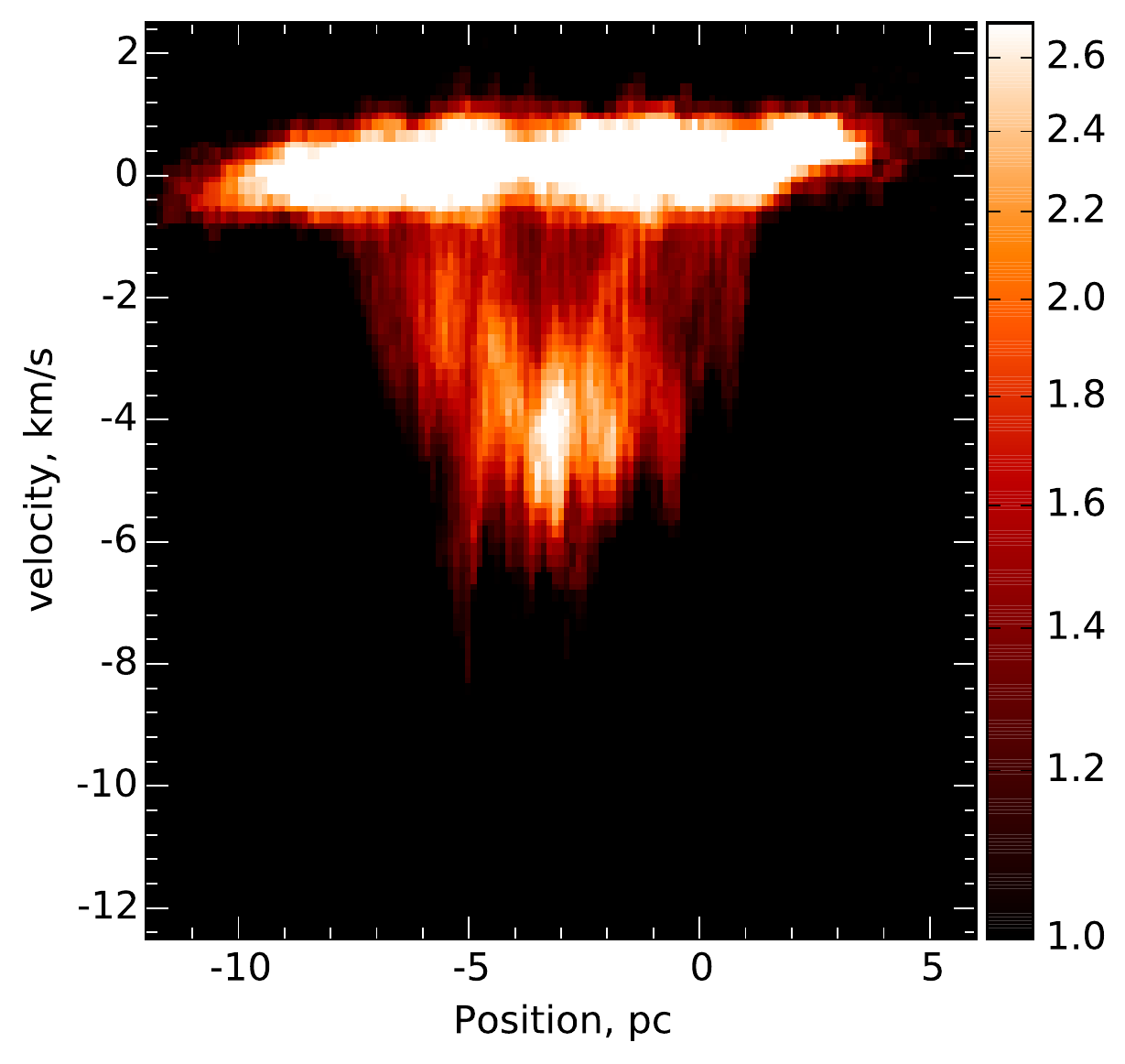}	
	\includegraphics[width=7.2cm]{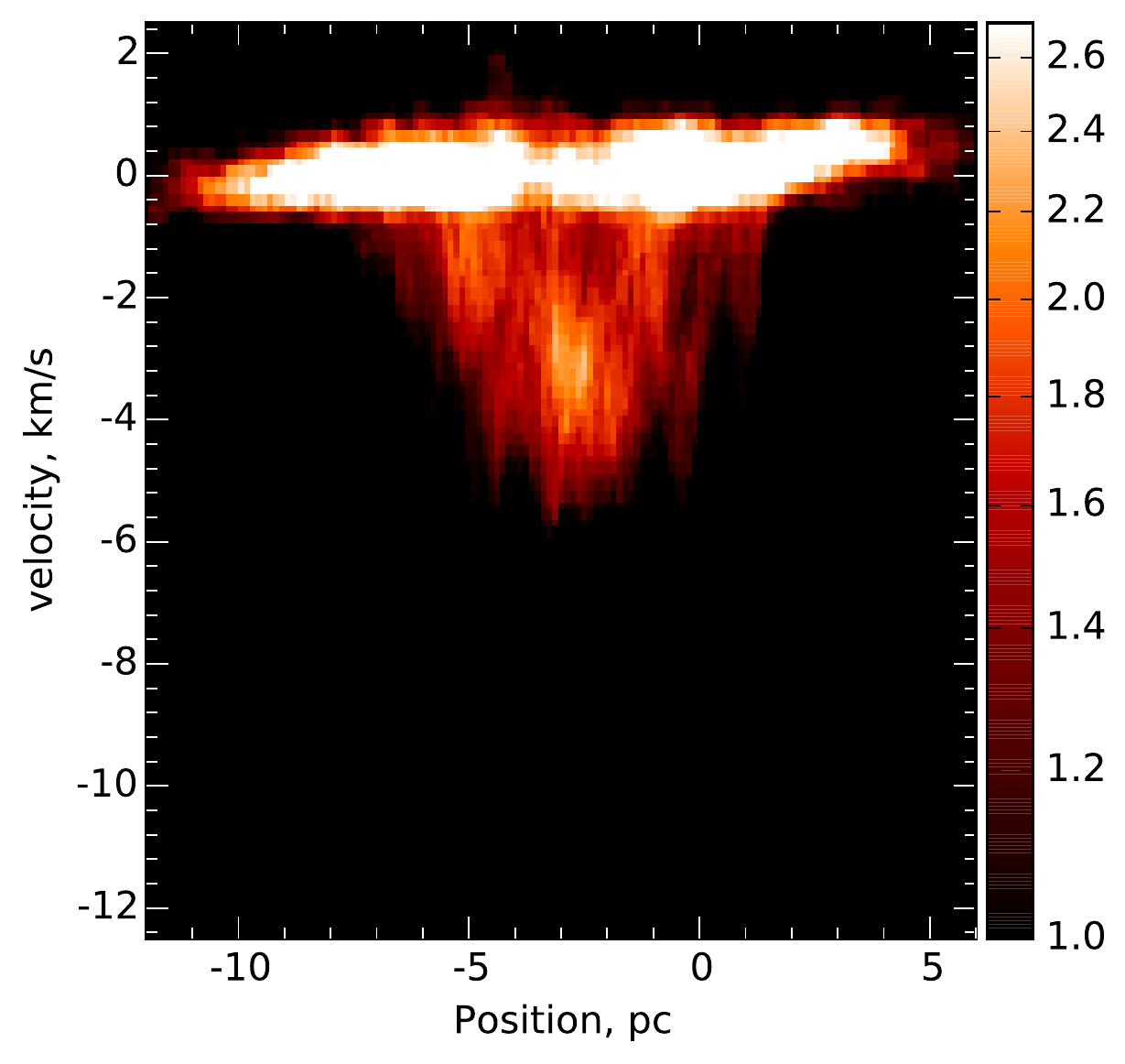}
	\caption{$^{12}$CO (J=$1\rightarrow0$) synthetic p--v diagrams of snapshots from the 10\,km\,s$^{-1}$ collision model from  Takahira et al (2014). The panels are for snapshots at 0.1 (top left), 1 (top right), 1.6 (bottom left)  and 2.6\,Myr (bottom right)..}
	\label{BB10}
\end{figure*}


\subsubsection{10km\,s$^{-1}$ collision model from Takahira et al {(2014)}}
We now apply our analytic arguments to the 10\,km\,s$^{-1}$ collision model from \cite{2014ApJ...792...63T}, the parameters of which are summarised in Table \ref{TakahiraModel}. Unlike in the previous model, this does not include any radiative feedback or star formation.  The disruption timescales for this model are shown in Figure \ref{allplot2}, where we assume an ionizing luminosity an order of magnitude lower than that in the previous model (10$^{47}$\,photons/s). In this collision between smaller clouds our analytic arguments predict that star formation as well as radiative and wind feedback will  be much more effective at disrupting the broad bridge - to the extent that they can readily dominate over the collision timescale. Recall that since the radiative disruption timescale scales as $R_s^{7/4}$ but only $N_{ly}^{-1/4}$, a reduction of the cloud size by a factor of 10 requires a reduction in the ionizing luminosity by a factor of $10^7$ to maintain the same disruption timescale. The simulations in  \cite{2014ApJ...792...63T} did not include star formation or feedback, so we expect the broad bridge to survive until the escape time.  Since the predicted escape timescale is smaller than the radiative one we do not expect radiative feedback would have been important to the broad bridge lifetime had it been included.

P--v diagrams for snapshots of this model at 0.1, 1, 1.6 and 2.6\,Myr are shown in Figure \ref{BB10}. There is similar evolution to the other collisional model, only with less small scale clumpy structure due to the lack of conversion of gas into stars, as well as a lack of feedback. The bottom right panel of Figure \ref{BB10} is 0.2\,Myr after the exit timescale predicts the broad bridge should be disrupted. Although some remnant of a secondary intensity peak remains it has been predominantly disrupted and is no longer so ``broad''. {The actual exit time in the simulation is about 2.8\,Myr, compared to our analytic estimate of 2.2\,Myr. This difference is consistent with our findings in section  2.3.1, where the actual collision velocity in the simulations was typically slightly lower than the analytic relations predict. This is due to lack of treatment of processes such as turbulent pressure and gravity.}

\section{Discussion}

\begin{figure}
	\hspace{-20pt}
	\includegraphics[width=9.6cm]{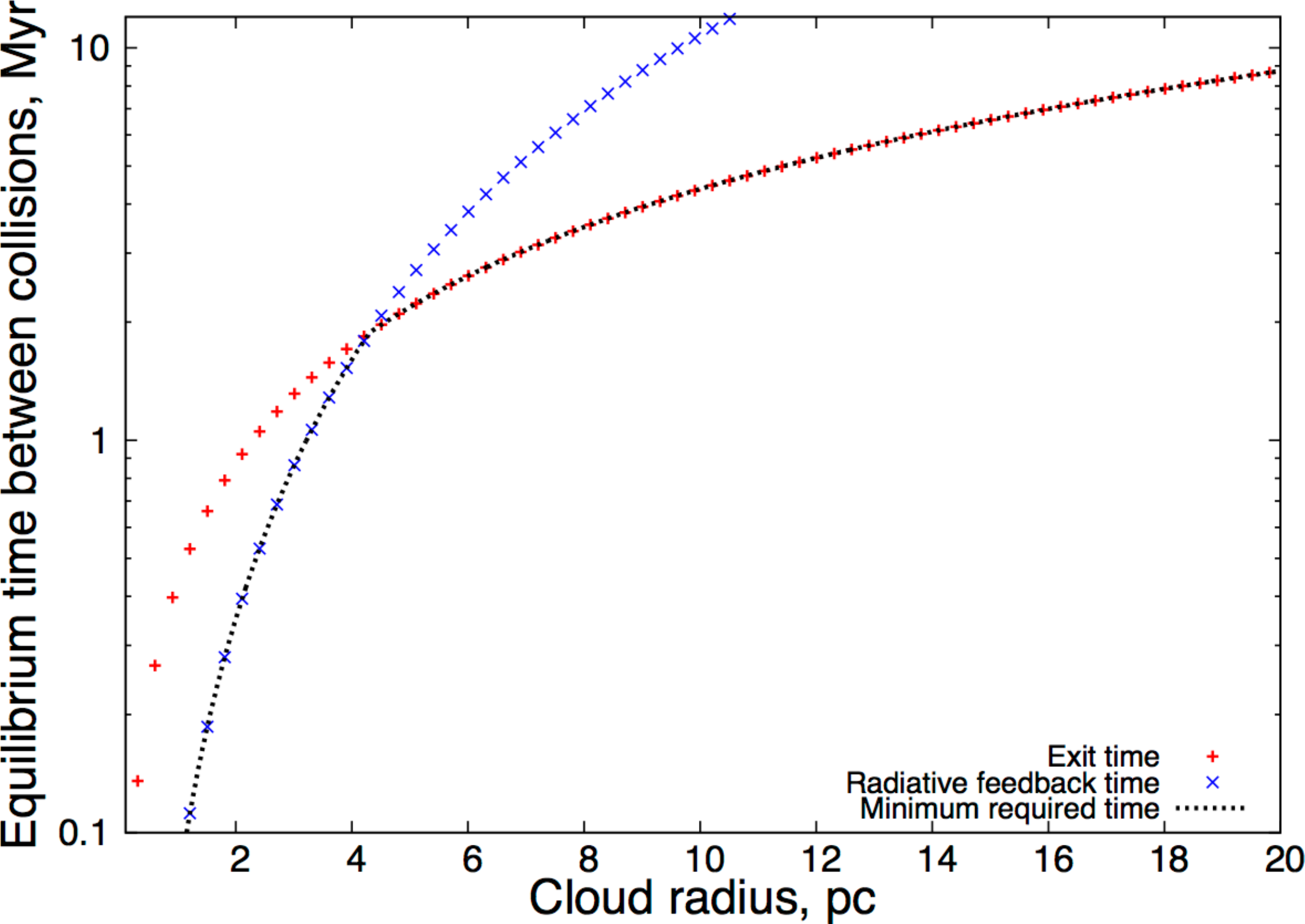}
	\caption{The time between cloud--cloud collisions as a function of cloud radius assuming that an equilibrium number of broad bridge features is maintained. This is for collision parameters discussed in section 4.1. }
	\label{equirate}
\end{figure}

\subsection{On the cloud--cloud collision rate}
We have developed tools for estimating broad bridge lifetimes which, where testable, seem to be consistent with our numerical models. In general we expect broad bridge lifetimes of around 1--5\,Myr. To date, seven broad bridge features have been identified in the Galaxy. Six of these are strong: M20 \citep{2011ApJ...738...46T, 2015MNRAS.450...10H}, RCW120 \citep{2015ApJ...806....7T}, NGC3603 \citep{2014ApJ...780...36F}, RCW38 \citep{2015arXiv150405391F}, NGC6334 and  NGC 6357. Westerlund 2 also shows a weak broad bridge feature \citep{2009ApJ...696L.115F}.  \cite{2015MNRAS.446.3608D} predict collisions on of order one every 10\,Myr for clouds of size 10\,pc or larger, which would be inconsistent with the amount of broad bridge features currently observed ({which are a lower limit, given that more may yet be observed}) if our typical broad bridge lifetimes are correct. 

However, in order for the number of broad bridges to not be extremely large (or zero), there should always be approximately an equilibrium number of broad bridges. In order to achieve a steady state, the rate at which broad bridges are formed must be approximately equal to the destruction rate. We can hence use our analytic arguments to estimate the equilibrium collision rate as a function of cloud size. {We do this in Figure \ref{equirate} for 10\,km\,s$^{-1}$ collisions. We assume that the larger cloud radius is twice that of the smaller.} {We assume that the smaller cloud has a density that scales as}
\begin{equation}
	n = \max(-2.1*\frac{R}{\textrm{pc}} + 54.75, 10.)
\end{equation}
{(which is based on the radii and densities of the clouds in the dynamical models discussed here) and that the larger cloud is half of this density.} {We also assume that the ionizing luminosity of the resulting stars varies with colliding cloud sizes as}
\begin{equation}
	N_{ly}  = \left(\frac{3\times10^{48}R_1}{14.45\,\textrm{pc}}\right)^{-2}\,\textrm{s}^{-1}
\end{equation}
{which is motivated by the Shima et al (in prep) simulation.}
We only consider the maximum required collision rate based on the exit timescale (equation \ref{texit}) and the radiative feedback timescale (equation \ref{big_t_r}). 

For clouds larger than about 5\,pc the exit timescales require collision rates consistent with those calculated by \cite{2015MNRAS.446.3608D}, but for smaller clouds \citep[not identified in the][estimates]{2015MNRAS.446.3608D}  the collision rates are expected to be much higher (this is required by both the exit and radiative feedback timescales). Of the systems observed to host broad bridges, NGC6334 and NGC6357 are approximately 10\,pc in size, however all others are less than 4\,pc. For the example of M20 which is about 2\,pc in size, from our equilibrium consideration we expect about 4 such collisions per Myr. This is in contrast to the  \cite{2015MNRAS.446.3608D}  general value of about 1 per 10\,Myr. It is therefore possible that high cloud--cloud collision rates, leading to massive star formation, are present in the Galaxy that are not contradictory to those rates expected from \cite{2015MNRAS.446.3608D}. {Higher collision rates for smaller clouds could be tested using new galactic models that are able to identify clouds below 10\,pc in size. If higher collision rates for smaller clouds are found, this would call for further detailed simulations of the collisional process. In particular, to understand the link between the properties of the colliding clouds (e.g. radius, density, turbulence, collision velocity) and any subsequent star formation (e.g. star formation efficiency and mass spectrum), which might be compared with observations of candidate collision sites.}

\section{Summary and conclusions}
We study the lifetimes of broad bridge features in p--v diagrams. We do so using a combination of analytic arguments and by post processing hydrodynamic and radiation hydrodynamic simulations of cloud--cloud collisions. We draw the following main conclusions from this work. \\

\noindent 1) We derived simple analytical tools to estimate broad bridge lifetimes. The lifetime of the broad bridge is constrained by either the collision timescale, the timescales for disruption by winds or radiative feedback and the timescale for the depletion of gas through star formation. \\

\noindent 2) The radiative feedback lifetime $t_R$ scales with the ionising flux as $t_R\propto N_{ly}^{-1/4}$ and the smaller cloud radius as $t_R\propto R_s^{7/4}$. This implies that to maintain a given radiative feedback disruption timescale for smaller clouds, the ionizing flux needs to be more drastically reduced. Due to this effect, although larger clouds have their broad bridge lifetimes determined by the collision timescale, for smaller clouds feedback can play a dominant role in determining the broad bridge lifetime. \\

\noindent 3) If we assume that there should be some constant number of broad bridges with time, then the disruption timescales should be roughly equal to the time between collisions at a given cloud size.  If this is the case our disruption timescales give collision rate estimates. We find that such estimates are readily consistent with the 1 per 10\,Myr collision rate estimates by \cite{2015MNRAS.446.3608D} {for clouds larger than about 6\,pc}. However for smaller clouds \citep[which][do not consider]{2015MNRAS.446.3608D}  we find much high collision rates, for example 4 per Myr for 2\,pc radius clouds. \\

\noindent 4) Given conclusion 3 and the fact that most high mass star formation sites where broad bridges have been identified are less than 4\,pc in size, we suggest that a large number of smaller scale cloud--cloud collisions could be triggering a substantial amount of high mass star formation.\\

\section*{Acknowledgements}
{We thank the referee for their insightful and thorough review, which led to the introduction of some interesting new aspects to the paper. }
TJH is funded by the STFC consolidated grant ST/K000985/1. EJT is funded by the MEXT grant for the Tenure Track System. Enzo simulations were carried out on the Cray XC30 at the Center for Computational Astrophysics (CfCA) of the National Astronomical Observatory of Japan. This research was supported by the DFG cluster of excellence `Origin and Structure of the Universe (JED)'. AH is supported  by   by JSPS KAKENHI Grant Number 15K05014. 

\bibliographystyle{mn2e}
\bibliography{molecular}

\appendix

\section{The thickness of the dense layer}
{Here we discuss the assumption that the slab thickness is less than the Str\"{o}mgren radius.
We begin by noting that this criterion is not essential and that so long as the layer is not much larger
than the Str\"{o}mgren radius (meaning that the H\textsc{ii} region will expand beyond the slab thickness
quite quickly) our results are unaffected.}  

{We can approximate the dense layer as a self-gravitating isothermal slab, for which the density distribution is given by}
\begin{equation}
	\rho_L(z) = \rho_L(0)\,\textrm{sech}^2\left(\frac{z}{H}\right)
\end{equation}
{where the scale height is }
\begin{equation}
	H = \left(\frac{c_s^2}{2 \pi G \rho_L}\right)^{1/2} = \left(\frac{c_s^4}{2 \pi G \rho_2 v_L^2}\right)^{1/2}
	\label{slabScale}
\end{equation}
{\citep[see e.g.][]{1999PASJ...51..625U, 2007pafd.book.....C}. We assume that the slab sound speed $c_s$ is the same as for all neutral gas (i.e. the same as that in the neutral colliding clouds). Equating this scale height to the Str\"{o}mgren radius (equation \ref{strom2}) we find that the assumption that the Str\"{o}mgren radius is larger than the slab width applies wherever}
\begin{equation}
	N_{ly} > \frac{ 2^{1/2}\alpha_B c_s^2v_L n_2^{1/2}}{3\pi^{1/2}(Gm_H)^{3/2}}.
\end{equation}
{Which, inserting standard values, can be expressed as }
\begin{equation}
	\left(\frac{N_{ly}}{10^{46}}\right) > 6.85 \left(\frac{v_L}{\textrm{km/s}}\right)\left(\frac{n_2}{\textrm{cm}^{-3}}\right)^{1/2}.
	\label{ThinCriterion}
\end{equation}
{Where we remind the reader that $v_L$ is the dense layer velocity and $n_2$ is the density in the larger cloud. For the example of the Shima et al (in prep) collision (discussed here in section 3.2.1) we find a predicted slab scale height of 0.05\,pc (so a slab width of about 0.1\,pc) and that equation \ref{ThinCriterion} is satisfied. We remind the reader that this analysis assumes that the ionising sources are centrally located. }

{A similar consideration can be made for winds by comparing equations \ref{slabScale} and \ref{rwind}, which tells us that the wind will break out of the slab after a time }
\begin{multline}
	\left(\frac{t}{\textrm{yr}}\right) > 50.67 
	\left(\frac{v_L}{\textrm{km/s}}\right)^{-1}
	\left(\frac{v_w}{\textrm{2000\,km/s}}\right)^{-1/2} \\ \times
	\left(\frac{n_2}{\textrm{cm}^{-3}}\right)^{-1/2} 
	\left(\frac{\dot{M}}{\textrm{M}_{\odot}\,\textrm{yr}^{-1}}\right)^{-1/2}.
\end{multline}

\bsp

\label{lastpage}

\end{document}